\documentstyle[12pt]{article}
\catcode`\@=11\relax
\newwrite\@unused
\def\typeout#1{{\let\protect\string\immediate\write\@unused{#1}}}
\def\figurepath{./}

\def\@nnil{\@nil}
\def\@empty{}
\def\@psdonoop#1\@@#2#3{}
\def\@psdo#1:=#2\do#3{\edef\@psdotmp{#2}\ifx\@psdotmp\@empty \else
    \expandafter\@psdoloop#2,\@nil,\@nil\@@#1{#3}\fi}
\def\@psdoloop#1,#2,#3\@@#4#5{\def#4{#1}\ifx #4\@nnil \else
       #5\def#4{#2}\ifx #4\@nnil \else#5\@ipsdoloop #3\@@#4{#5}\fi\fi}
\def\@ipsdoloop#1,#2\@@#3#4{\def#3{#1}\ifx #3\@nnil 
       \let\@nextwhile=\@psdonoop \else
      #4\relax\let\@nextwhile=\@ipsdoloop\fi\@nextwhile#2\@@#3{#4}}
\def\@tpsdo#1:=#2\do#3{\xdef\@psdotmp{#2}\ifx\@psdotmp\@empty \else
    \@tpsdoloop#2\@nil\@nil\@@#1{#3}\fi}
\def\@tpsdoloop#1#2\@@#3#4{\def#3{#1}\ifx #3\@nnil 
       \let\@nextwhile=\@psdonoop \else
      #4\relax\let\@nextwhile=\@tpsdoloop\fi\@nextwhile#2\@@#3{#4}}
\def\psdraft{
	\def\@psdraft{0}
}
\def\psfull{
	\def\@psdraft{100}
}
\psfull
\newif\if@prologfile
\newif\if@postlogfile
\newif\if@noisy
\def\pssilent{
	\@noisyfalse
}
\def\psnoisy{
	\@noisytrue
}
\psnoisy
\newif\if@bbllx
\newif\if@bblly
\newif\if@bburx
\newif\if@bbury
\newif\if@height
\newif\if@width
\newif\if@rheight
\newif\if@rwidth
\newif\if@clip
\newif\if@verbose
\def\@p@@sclip#1{\@cliptrue}

\def\@p@@sfile#1{\def\@p@sfile{null}%
	        \openin1=#1
		\ifeof1\closein1%
		       \openin1=\figurepath#1
			\ifeof1\typeout{Error, File #1 not found}
			\else\closein1
			    \edef\@p@sfile{\figurepath#1}%
                        \fi%
		 \else\closein1%
		       \def\@p@sfile{#1}%
		 \fi}
\def\@p@@sfigure#1{\def\@p@sfile{null}%
	        \openin1=#1
		\ifeof1\closein1%
		       \openin1=\figurepath#1
			\ifeof1\typeout{Error, File #1 not found}
			\else\closein1
			    \def\@p@sfile{\figurepath#1}%
                        \fi%
		 \else\closein1%
		       \def\@p@sfile{#1}%
		 \fi}

\def\@p@@sbbllx#1{
		\@bbllxtrue
		\dimen100=#1
		\edef\@p@sbbllx{\number\dimen100}
}
\def\@p@@sbblly#1{
		\@bbllytrue
		\dimen100=#1
		\edef\@p@sbblly{\number\dimen100}
}
\def\@p@@sbburx#1{
		\@bburxtrue
		\dimen100=#1
		\edef\@p@sbburx{\number\dimen100}
}
\def\@p@@sbbury#1{
		\@bburytrue
		\dimen100=#1
		\edef\@p@sbbury{\number\dimen100}
}
\def\@p@@sheight#1{
		\@heighttrue
		\dimen100=#1
   		\edef\@p@sheight{\number\dimen100}
}
\def\@p@@swidth#1{
		\@widthtrue
		\dimen100=#1
		\edef\@p@swidth{\number\dimen100}
}
\def\@p@@srheight#1{
		\@rheighttrue
		\dimen100=#1
		\edef\@p@srheight{\number\dimen100}
}
\def\@p@@srwidth#1{
		\@rwidthtrue
		\dimen100=#1
		\edef\@p@srwidth{\number\dimen100}
}
\def\@p@@ssilent#1{ 
		\@verbosefalse
}
\def\@p@@sprolog#1{\@prologfiletrue\def\@prologfileval{#1}}
\def\@p@@spostlog#1{\@postlogfiletrue\def\@postlogfileval{#1}}
\def\@cs@name#1{\csname #1\endcsname}
\def\@setparms#1=#2,{\@cs@name{@p@@s#1}{#2}}
\def\ps@init@parms{
		\@bbllxfalse \@bbllyfalse
		\@bburxfalse \@bburyfalse
		\@heightfalse \@widthfalse
		\@rheightfalse \@rwidthfalse
		\def\@p@sbbllx{}\def\@p@sbblly{}
		\def\@p@sbburx{}\def\@p@sbbury{}
		\def\@p@sheight{}\def\@p@swidth{}
		\def\@p@srheight{}\def\@p@srwidth{}
		\def\@p@sfile{}
		\def\@p@scost{10}
		\def\@sc{}
		\@prologfilefalse
		\@postlogfilefalse
		\@clipfalse
		\if@noisy
			\@verbosetrue
		\else
			\@verbosefalse
		\fi
}
\def\parse@ps@parms#1{
	 	\@psdo\@psfiga:=#1\do
		   {\expandafter\@setparms\@psfiga,}}
\newif\ifno@bb
\newif\ifnot@eof
\newread\ps@stream
\def\bb@missing{
	\if@verbose{
		\typeout{psfig: searching \@p@sfile \space  for bounding box}
	}\fi
	\openin\ps@stream=\@p@sfile
	\no@bbtrue
	\not@eoftrue
	\catcode`\%=12
	\loop
		\read\ps@stream to \line@in
		\global\toks200=\expandafter{\line@in}
		\ifeof\ps@stream \not@eoffalse \fi
		\@bbtest{\toks200}
		\if@bbmatch\not@eoffalse\expandafter\bb@cull\the\toks200\fi
	\ifnot@eof \repeat
	\catcode`\%=14
}	
\catcode`\%=12
\newif\if@bbmatch
\def\@bbtest#1{\expandafter\@a@\the#1
\long\def\@a@#1
\long\def\bb@cull#1 #2 #3 #4 #5 {
	\dimen100=#2 bp\edef\@p@sbbllx{\number\dimen100}
	\dimen100=#3 bp\edef\@p@sbblly{\number\dimen100}
	\dimen100=#4 bp\edef\@p@sbburx{\number\dimen100}
	\dimen100=#5 bp\edef\@p@sbbury{\number\dimen100}
	\no@bbfalse
}
\catcode`\%=14
\def\compute@bb{
		\no@bbfalse
		\if@bbllx \else \no@bbtrue \fi
		\if@bblly \else \no@bbtrue \fi
		\if@bburx \else \no@bbtrue \fi
		\if@bbury \else \no@bbtrue \fi
		\ifno@bb \bb@missing \fi
		\ifno@bb \typeout{FATAL ERROR: no bb supplied or found}
			\no-bb-error
		\fi
		\count203=\@p@sbburx
		\count204=\@p@sbbury
		\advance\count203 by -\@p@sbbllx
		\advance\count204 by -\@p@sbblly
		\edef\@bbw{\number\count203}
		\edef\@bbh{\number\count204}
}
\def\in@hundreds#1#2#3{\count240=#2 \count241=#3
		     \count100=\count240	
		     \divide\count100 by \count241
		     \count101=\count100
		     \multiply\count101 by \count241
		     \advance\count240 by -\count101
		     \multiply\count240 by 10
		     \count101=\count240	
		     \divide\count101 by \count241
		     \count102=\count101
		     \multiply\count102 by \count241
		     \advance\count240 by -\count102
		     \multiply\count240 by 10
		     \count102=\count240	
		     \divide\count102 by \count241
		     \count200=#1\count205=0
		     \count201=\count200
			\multiply\count201 by \count100
		 	\advance\count205 by \count201
		     \count201=\count200
			\divide\count201 by 10
			\multiply\count201 by \count101
			\advance\count205 by \count201
		     \count201=\count200
			\divide\count201 by 100
			\multiply\count201 by \count102
			\advance\count205 by \count201
		     \edef\@result{\number\count205}
}
\def\compute@wfromh{
		\in@hundreds{\@p@sheight}{\@bbw}{\@bbh}
		\edef\@p@swidth{\@result}
}
\def\compute@hfromw{
		\in@hundreds{\@p@swidth}{\@bbh}{\@bbw}
		\edef\@p@sheight{\@result}
}
\def\compute@handw{
		\if@height 
			\if@width
			\else
				\compute@wfromh
			\fi
		\else 
			\if@width
				\compute@hfromw
			\else
				\edef\@p@sheight{\@bbh}
				\edef\@p@swidth{\@bbw}
			\fi
		\fi
}
\def\compute@resv{
		\if@rheight \else \edef\@p@srheight{\@p@sheight} \fi
		\if@rwidth \else \edef\@p@srwidth{\@p@swidth} \fi
}
\def\compute@sizes{
	\compute@bb
	\compute@handw
	\compute@resv
}
\def\psfig#1{\vbox {
	\ps@init@parms
	\parse@ps@parms{#1}
	\compute@sizes
	\ifnum\@p@scost<\@psdraft{
		\if@verbose{
			\typeout{psfig: including \@p@sfile \space }
		}\fi
		\special{ps::[begin] 	\@p@swidth \space \@p@sheight \space
				\@p@sbbllx \space \@p@sbblly \space
				\@p@sbburx \space \@p@sbbury \space
				startTexFig \space }
		\if@clip{
			\if@verbose{
				\typeout{(clip)}
			}\fi
			\special{ps:: doclip \space }
		}\fi
		\if@prologfile
		    \special{ps: plotfile \@prologfileval \space } \fi
		\special{ps: plotfile \@p@sfile \space }
		\if@postlogfile
		    \special{ps: plotfile \@postlogfileval \space } \fi
		\special{ps::[end] endTexFig \space }
		\vbox to \@p@srheight true sp{
			\hbox to \@p@srwidth true sp{
				\hss
			}
		\vss
		}
	}\else{
		\vbox to \@p@srheight true sp{
		\vss
			\hbox to \@p@srwidth true sp{
				\hss
				\if@verbose{
					\@p@sfile
				}\fi
				\hss
			}
		\vss
		}
	}\fi
}}
\def\psglobal{\typeout{psfig: PSGLOBAL is OBSOLETE; use psprint -m instead}}
\catcode`\@=12\relax
\begin{document}
\parindent=1cm
\parskip=5pt
\baselineskip=22pt
\renewcommand{\thefootnote}{\fnsymbol{footnote}}
\begin{titlepage}
~~~~~~~~~~~~~~~\\
\vspace{-50pt}
\baselineskip=15pt
\begin{flushright}
CU-TP-792\\
CAL-621\\
\end{flushright}
\baselineskip=20pt
\begin{center}
{\bf KINETICS OF ELECTRON--POSITRON PAIR PLASMAS USING AN
ADAPTIVE MONTE CARLO METHOD}\footnote{To appear in the Astrophysical Journal}\\
\smallskip
Ravi P. Pilla\footnote{ravi@cuphyb.phys.columbia.edu} and Jacob Shaham\footnote
{Dr. Shaham passed away during the course of this work}\\
Department of Physics, Columbia University\\
538 West $120^{\rm th}$ Street, New York, NY 10027, USA\\
\end{center}
\vfil
\centerline{\bf ABSTRACT}

A new algorithm for implementing the adaptive Monte Carlo method is given. 
It is used to solve the Boltzmann equations that describe the time 
evolution of a nonequilibrium electron--positron pair plasma 
containing high-energy photons. These are coupled nonlinear 
integro-differential equations. 
The collision kernels for the photons as well as pairs are evaluated
for Compton scattering, pair annihilation and  creation, bremsstrahlung, and
Coulomb collisions.
They are given as multidimensional
integrals which are valid for all energies. For an homogeneous and
isotropic plasma with no particle escape,
the equilibrium solution is expressed analytically 
in terms of the initial conditions. For two specific cases, for which
the photon and the pair spectra are initially constant or have a power law
distribution within the given limits, the time evolution of the plasma is 
analyzed using the new method. The final spectra are found to be in a
good agreement with the analytical solutions. 
The new algorithm is faster
than the Monte Carlo scheme based on uniform sampling and more flexible
than the numerical methods used in the past, which do not involve
Monte Carlo sampling. 
It is also found to be very stable.
Some astrophysical applications of this technique are discussed.

\parindent=0cm
{\it Subject headings:} $\gamma$-rays: theory -- methods:
Monte Carlo -- plasmas: relativistic kinetic theory -- 
radiation mechanisms: nonthermal
-- X-rays: theory
\parindent=1cm
\vfil
\end{titlepage}
%
%
%
%
%
\def\theequation{{\arabic{section}.\arabic{equation}}}
\def\reset{\setcounter{equation}{0}}
\def\mc{Monte Carlo }
\def\pp{pair plasma }
\def\nepp{non-equilibrium pair plasma }
\def\nm{n_{-}}
\def\np{n_{+}}
\def\ne{n_{e}}
\def\nph{n_{\gamma}}
\def\mysection#1{\vskip 1pt\vspace{-1pt} \centerline{\bf #1}\vspace{-3pt}}
\def\ee{\varepsilon}
\def\fkin{{\cal F}}
\def\bb{\mbox{\boldmath $\beta$}}  
\def\brel{\beta_{\rm rel}}
\def\beq{\begin{equation}}
\def\eeq{\end{equation}}
\newcommand{\mysubsec}[1]{\vspace{-13pt}\begin{flushleft}
{\sc #1}\end{flushleft}\vspace{-12pt}}
\newcommand{\re}{r_{\rm e}}
\def\brem{bremsstrahlung}
\def\deby{\lambda_{\rm D}}
%
%
%
%
%
%
%
\baselineskip=20pt
\def\mc{Monte-Carlo }
\def\pp{pair plasma }
\def\nepp{non-equilibrium pair plasma }
\def\nm{n_{-}}
\def\np{n_{+}}
\def\ne{n_{e}}
\def\nph{n_{\gamma}}
\centerline{\bf 1 Introduction}
\par
Nonthermal emission of high-energy radiation from a variety of  compact
astrophysical objects e.g., $\gamma$-ray-burst sources  (M\'esz\'aros 
\& Rees 1993a,b), pulsars (Chen \& Ruderman 1993), active galactic 
nuclei (AGN; Lightman \& Zdziarski 1987; Svensson 1994; and Padovani 1996), 
and jets in the
AGN (Sikora 1994) seem to indicate the presence of a relativistic 
electron-positron pair plasma in the dense radiation fields of those
sources. Such plasmas may exist also in the accretion disc coronas of 
the Galactic X-ray binaries (Sunyaev et al. 1992), the ergo-spheres of
Kerr black holes (Piran \& Shaham 1977), and the black-hole
accretion discs (Tanaka \& Kusunose 1985; Gunnlaugur \& Svensson
1992). It is conceivable that the pair plasma in some of these
sources is in thermodynamic equilibrium with itself and probably
in equilibrium with the radiation. However, it is more plausible that
many of them may consist of nonequilibrium pair plasmas
(Coppi \& Blandford 1990 -- CB90 henceforth; Zdziarski 1988 and  1989). 
Many of the previous papers on this topic have concentrated on the the 
properties of a relativistic pair plasma in thermal equilibrium (e.g.,
Bisnovatyi-Kogan, Zel'dovich, \& Sunyaev 1971; Lightman \& Band 1981;
Lightman 1981 and 1982; Svensson 1982b -- henceforth S82b; and Zdziarski 1985).
Examples of the time evolution of a thermal pair plasma, taking into account
the finite-medium radiative transfer effects can be found in
Guilbert \& Stepney (1985) and Kusunose (1987).
There are not many papers that deal with the evolution of a 
nonequilibrium pair plasma in detail; some examples can be found in
Lightman \& Zdziarski (1987), Svensson (1987),
Zdziarski, Coppi, \& Lamb (1990), and Coppi (1992 -- C92 henceforth). 

These investigations are generally based on the  Monte Carlo (MC) methods or
on solving the Boltzmann equations (kinetic theory approach).
In a simple MC  method based on uniform sampling (Pozdnyakov, Sobol' \& 
Sunyaev 1977), individual particles are followed as they undergo interactions
in the source. In this method, it is usually easy to take into account the 
spatial inhomogeneities and radiative transfer effects well. But it typically 
suffers from relatively poor photon statistics at higher energies and does
not lend itself to time-evolution calculations involving broad-band spectra.
For examples of such MC simulations, see Novikov \& Stern (1986).
In the kinetic theory approach, the system is represented by 
the photon and particle distribution functions which are discretized in energy
as well as the spatial coordinates and the time evolution is determined by
solving the Boltzmann equations numerically. 
In general, it is very difficult to solve 
the resulting integro-differential equations. Moreover, they are usually 
``stiff'' (i.e., there are very different time-scales in the problem).
The principal advantage of this approach is that it gives good photon 
statistics at higher energies. Some
examples of the kinetic theory approach can be found in C92,
Ghisellini (1987), Svensson (1987), and Fabian {\it et al.}
(1986). 

There have been some  attempts to improve the photon statistics
in the conventional MC schemes which go by the name phase-space density
(PSD) array representation. In this approach, the system is represented 
by the discretized distribution functions (as in the kinetic theory approach)
but the particle or photon transitions between the energy bins is simulated
using the MC method and the interaction between the spatial cells is 
modeled with the aid of the escape probabilities. So far this approach has 
been used to model only homogeneous and spherically symmetric systems (e.g.,
Stern 1985). Another recent variant of the MC method is based on the
large-particle (LP) representation, which is described in detail 
by Stern {\it et al.}
(1995). In this scheme, the system is represented by an array of ``large
particles'', each of which corresponds to a group of real particles sharing
the same physical parameters (i.e., particle type, position, momentum, 
and energy). It is more flexible than the PSD approach in the sense that 
each LP is tagged with a statistical weight, which is proportional to the 
number of real particles represented by that LP. 
For example, this weight can be assigned based on the total energy carried
by each LP. In many nonequilibrium systems of interest in astrophysics, 
the number of particles in the low-energy range is typically several orders
of magnitude larger than that of the particles in the high-energy range. 
Therefore, the efficiency of the method may be improved by assigning lower
statistical weight to the low-energy LPs. Intuitively this approach makes
sense but there is no general proof for its validity or  effectiveness 
(except for the  numerical experiments presented by Stern {\it et al.} 1995).
Besides, the statistical weights are rather {\it ad hoc.} 

From the preceding discussion, it is clear that the  main problem in the 
analysis of nonequilibrium pair 
plasmas is the computational difficulty. The principal aim of this paper is
to present a new method for  solving the kinetic equations based on an
adaptive MC sampling scheme. 
It is faster than the conventional MC method (based
on uniform sampling) and is more flexible (and in some cases, faster) 
than the numerical methods previously used.  Our method 
resembles the LP method described above, in the usage of the statistical
weights, but it is much more rigorous. Moreover, it can accommodate 
anisotropic distributions with greater ease.
 
In a relativistic
plasma containing arbitrary densities of pairs and the
high energy photons, the collision cross sections for various
microscopic processes depend on the energy. One cannot
use, for example, the simple Thomson cross section as one can do in the 
nonrelativistic case. In addition  there is a creation and 
annihilation of the pairs and photons that  alter the densities.
Therefore we have to follow the time evolution of the 
number density as well as the spectrum of each species. 
Besides, the problem is inherently nonlinear due to the form of the 
collision kernels in the Boltzmann equations. 
It is possible to write all the collision kernels as 
multidimensional integrals. This reduces the problem of solving the
coupled Boltzmann equations for the photons and the pairs into 
a purely computational task of evaluating many of these integrals,
after each time step, quickly and efficiently. This way of formulating
the problem of kinetic theory is more flexible in accommodating any kind
of distribution functions. We have developed a new algorithm,
based on Monte Carlo sampling,  for computing such integrals.
The functional form of the integrands
is not assumed a priori. Also, no constraint is placed on the shape of the 
integration region. Usually such integrals are evaluated either numerically
(by using an equally spaced discrete grid) or through a Monte Carlo 
sampling technique. 
In order to make the former method more efficient, 
we have to choose the shape of the discrete mesh depending on the form of 
the integrand. This takes away the flexibility from the method (i.e., 
the algorithm will depend on the form of the integrand).
The latter method, based on uniform sampling throughout the 
integration region, is widely used in astrophysics.
It is possible to speed up the computation 
in this method, by sampling selectively i.e.,
sampling more frequently in those domains where the integrand is larger.
This scheme is known as the 
importance sampling method or the  adaptive Monte Carlo
method. There is an algorithm, originally due to Lepage (1978), which
implements this. However it is not well suited for the type of integrals that 
arise in the present context. We have developed a new algorithm to implement
the adaptive Monte Carlo method which is very efficient (see below).

In the next section we define various quantities, explain the basic pair 
plasma model we use, 
and write down the general kinetic equations. In sections $3$ and 
$4$ we give the integral expressions for various collision kernels, that are 
valid  for all energies. These collision integrals are cast in a form that is
well suited for the Monte Carlo integration. 
In section $5$ we describe how we integrate the 
Boltzmann equations numerically. There we explain the adaptive Monte Carlo 
algorithm we use. In section $6$ we express the final equilibrium state of 
an homogeneous and isotropic plasma (with no escape of particles or photons) 
analytically in terms of the initial conditions. Then we apply our 
time-evolution code to  two specific examples of nonequilibrium configurations 
and compare the final results with the corresponding analytical solutions. 
These examples serve as a test for the overall formalism presented in this
paper. Finally, in section $7$, we
summarize this work and discuss some astrophysical applications.
\medskip
%
%
%
%
%
%
\mysection{2 Model, definitions, and the notation}
\smallskip
\setcounter{section}{2}
\reset

We consider a neutral, stationary, and unmagnetized pair plasma
which is nonthermal (i.e., not in equilibrium). 
We assume that the plasma is homogeneous
and isotropic. If the plasma is in a moving source
we must interpret all the physical quantities given below as the comoving-frame
quantities. The number densities (i.e., the number of particles per
unit volume) of the electrons, positrons, photons, and protons are
given by: $\nm, \np, \nph$, and $n_{\rm p}$, respectively 
$(n_{-}=n_{+}+n_{\rm p})$. 
Throughout this paper we express the momentum and energy in 
units of $mc$ and $mc^{2}$, respectively. Here $m$ is the electron
rest mass and $c$ is the speed of light in free space. Therefore the
momenta and the energies of the particles, as well as the 
photons, are represented
by dimensionless numbers everywhere. For the models we consider here
the protons can be assumed to be at rest.
We assume that the state of the plasma is completely described by the 
Lorentz invariant distribution functions
$f_{\pm}(x,p)$ and $f_{\gamma}(x,p)$, 
for positrons, electrons, and photons, respectively. Here $x,p$ represent
the position and the momentum four-vectors, respectively and
$\bf x,p$ represent the
corresponding three-vectors. Our choice of the metric is 
such that $p^{2}=1$ for electrons. In the case of photons we have 
$p=\ee (1,{\bf k})$, where $\ee$ is the photon energy and 
$\bf k$ is a unit vector in the direction of its three-momentum.
Similarly, $p = \gamma(1,\mbox{\boldmath $\beta$})$ for the pairs. 
Here $\gamma$ is the Lorentz factor and {\boldmath $\beta$}
is the velocity in units of c. We denote the magnitude of
{\boldmath $\beta$} by $\beta$.
The number density of the particles of type $i$ with a  momentum $\bf p$
is given by $f_{i}\,d^{3}{\bf p}$. 
We define the total densities of various species to be $n_{i}=
\int f_{i}
({\bf p})\,d^{3}{\bf p}$, where the integration extends over all values 
of the momenta. Because of the isotropy, we have $d^{3}{\bf p}=4\pi\ee^{2}d 
\ee$  in the case of  photons and $d^{3}{\bf p}=4\pi\beta\,
\gamma^{2}\,d\gamma$ for the pairs.

Since we assume that the plasma is homogeneous and isotropic, various
distribution functions depend only on time  and the energy (or the 
magnitude of the momentum). We define the spectral
functions for photons, positrons, and electrons to be
\beq
F_{\gamma}(\ee)= \frac{4 \pi \ee^{2}}{\nph}f_{\gamma}(\ee)\quad{\rm and}\quad
F_{\pm}(\gamma)=\frac{4 \pi\,\beta\,\gamma^{2}}{n_{\pm}}f_{\pm}(\gamma),
\label{spectral-F}
\eeq
respectively. The time dependence of these functions is not shown explicitly.
The spectral functions are normalized so that
\beq
\int_{0}^{\infty}d\ee F_{\gamma}(\ee)=1
\quad{\rm and}\quad\int_{1}^{\infty}d\gamma F_{\pm}(\gamma)=1.
\eeq
We see that the number of photons of  energy $\ee$ per unit
volume and unit energy is given by $\nph F_{\gamma}(\ee)$.
We will assume that the electrons and the positrons have the same 
spectral functions i.e., $F_{-}(\gamma)=F_{+}(\gamma)$
for all values of $\gamma$, which we denote by $F_{e}(\gamma)$.

The equilibrium spectral functions, which are
independent of time, are given by
\beq
F_{\gamma}(\ee)=\frac{1}{2 \zeta(3) \Theta^{3}}\: \frac{\ee^{2}}
{{\rm exp}(\ee/\Theta)-1}
\label{planck}
\eeq
and
\beq
F_{e}(\gamma)=\frac{1}{\Theta K_{2}(1/\Theta)}\:\beta\,\gamma^{2}\;
{\rm exp}(-\gamma/\Theta).
\label{maxwell}
\eeq
Equation (\ref{planck}) comes from the Planck function for the photons, where
$\zeta$ is
the Riemann zeta function and $\zeta(3)\cong1.202$.
In that equation we have used the equilibrium
density of photons
\beq
\nph=16\pi\zeta(3)\:\left(\frac{mc}{h}\Theta\right)^{3},
\label{eqbm-nph}
\eeq
where $h$ is the Planck's constant.
Equation (\ref{maxwell}) is the relativistic Maxwell-Boltzmann
distribution for electrons and $K_{2}$ is the second order modified 
Bessel function of the second kind. In all these equations
$\Theta = k_{B}T/mc^2$ is the dimensionless temperature of the 
plasma, where $T$ is the temperature and $k_{B}$ is the Boltzmann constant.
\par To study the time evolution of this system we 
should proceed from the relativistic Boltzmann equations for 
the pairs and photons. In the latter case it is the same as the
radiative transfer equation. The Boltzmann equation 
(see e.g., de Groot, van Leeuwen, \& van Weert 1980) for the   
particles of type $i$, described by $f_{i}$, which takes into
account the collisions with the particles of  
type $j$, described by $f_{j}$, is given by 
\beq
p^{\mu}\partial_{\mu}\,f_{i}(x,p)=\sum_{j}\int\frac{d^{3}{\bf q}}{q^{0}}
d\Omega^{\prime}
\,\left[f_{i}(x,p^{\prime})f_{j}(x,q^{\prime})-f_{i}(x,p)f_{j}(x,q)
\right]\,F\sigma_{ij}.
\label{main-boltzmann}
\eeq
Here $\partial_{\mu}$ is the partial derivative with respect to 
$x^{\mu}$ and the summation over $\mu$ is implied. The summation for
$j$ extends over all relevant processes. Here $q^{0}$ is the 
energy component of the 4-vector $q$. Using the initial and the final 
momenta to designate the particles, the collision processes can be
represented as $p+q\leftrightarrow p^{\prime}+q^{\prime}$. The 
solid angle around one of the outgoing particles is $d\Omega^{\prime}$.
Finally,
$\sigma_{ij}$ is the cross section for the process and F is the 
invariant flux factor.
It is necessary to remark that in the present form, the 
above equation cannot account for the quantum mechanical Bose enhancement and 
Fermi blocking effects, respectively for the  photons and pairs. In order
to do so, we need to take into account the
particle occupation numbers in the phase space.
For photons, this is given by
\beq
g_{\gamma}(\ee)=\frac{1}{2}\left(\frac{h}{mc}\right)^{3}f_{\gamma}(\ee)=
\left(\frac{h}{mc}\right)^{3}\frac{\nph F_{\gamma}(\ee)}{8\pi\ee^{2}},
\label{occupation}
\eeq
which in the equilibrium case  reduces to $1/[{\rm exp}(\ee/\Theta)-1]$, as
expected. If we are considering a process in which two particles
of momenta $p$ and $q$ produce a photon of momentum $p^{\prime}$, then
we should make the replacement $f_{i}(p)f_{j}(q)\rightarrow f_{i}(p)f_{j}(q)
[1+g_{\gamma}(p^{\prime})]$ in the Boltzmann equation. 
These effects play a significant role
only when $g_{\gamma}\simeq 1$ or $\nph\,\ee^{-2}F_{\gamma}(\ee)\simeq 1.76
\times 10^{30} {\rm cm}^{-3}$. For the densities and the 
energies of interest here,
these quantum mechanical effects can be neglected. An analogous remark 
applies to the case of the pairs. Such induced effects in a relativistic 
thermal plasma at high temperatures and densities have been considered 
by many authors in the past (e.g., Ramaty, McKinley, and Jones 1982).  

\par The Boltzmann equations reduce to simple rate equations in the comoving
frame as a result of the homogeneity and isotropy of the plasma.  We denote 
the comoving time coordinate by $t$. The rate equations are given by
\beq
\frac{\partial}{\partial t}f_{i}=
\sum_{q}\:\left[\eta_{i}-f_{i}\:\chi_{i}\right]_{q},
\label{boltzmann}
\eeq
where $i$ stands for either photons or electrons and  $q$ 
labels the binary collision process (Compton scattering, pair 
processes, bremsstrahlung, or Coulomb collisions). 
The summation runs over
all those processes that involve a particle of type $i$ among the
products of the collision. Here $\eta_{i}$ is the emission 
coefficient for the production  of a particle
of type $i$ with momentum $p$ (or scattering of such a particle
into that final momentum state) and  $\chi_{i}$ is the corresponding absorption
coefficient. Notice that $f_{i},\,\eta_{i},\,{\rm and}\,\chi_{i}$
depend only on the energy of the particles and time. 
In order to obtain the 
collision kernels, $\eta_{i}$
and $\chi_{i}$, we require the binary reaction rates in a relativistic
plasma (e.g., de Groot, van Leeuwen, \& van Weert 1980; Baring 1987a). 
Using the appropriate reaction rates we can write
\beq
\eta_{i}\,(p)=\sum_{l,m}\:\frac{c}{1+\delta_{lm}}\,\int_{U}
\:dF_{lm}\,\fkin_{lm}\,
\frac{d\sigma_{lm}}{dP}\,,
\label{eta}
\eeq
where $\delta_{lm}=1$ for identical colliding particles (i.e., $l=m$) and is
zero otherwise. The summation in this equation is over those incident states
(labeled by $l$ and $m$) which result in a final state labeled by $i$.
Furthermore, $d\sigma_{lm}/dP$
is the differential cross section for the process whereas $dP$ is a 
shorthand for $d^{3}{\bf p}$ which is defined above.
The four-momenta of the colliding particles are given by $p_{k}=(p_{k}^{0},
{\bf p}_{k})$ (for $k=l,m$) and
the four-momentum of one of the outgoing particles is $p$. 
The product of the phase-space densities of the
colliding particles $dF_{lm}$ is given by
\beq
dF_{lm}=\prod_{j=l,m}\,f_{j}(p_{j})\,d^{3}{\bf p}_{j}.
\label{dF12}
\eeq
We have
$d^{3}{\bf p}_{l}= \ee^{2}_{l}d\ee_{l} d\Omega_{l}$ for the photons 
and $d^{3}{\bf p}_{l}=
\beta_{l}\,\gamma_{l}^{2}\,d\Omega_{l}\,d\gamma_{l}$ for the pairs. The
kinematic factor $\,\fkin_{lm}$ for binary collisions 
(see e.g., Landau \& Lifshitz 1975) is given by
\beq
\fkin_{lm}=(u_{l}\cdot u_{m})\,\brel(p_{l},p_{m}),
\label{fkin}
\eeq
where $u_{l}=p_{l}/p^{0}_{l}$ and $\brel$ is the 
relative velocity of the colliding particles in units of $c$.
If at least one of the colliding particles is a photon 
we will have $\brel=1$. Otherwise 
\beq
\brel\,(p_{l},p_{m})=\frac{\left[(\bb_{l} - \bb_{m})^{2}-
(\bb_{l}\times \bb_{m})^{2}\right]^{1/2}}{1-\bb_{l}\cdot \bb_{m}}.
\label{rel-beta}
\eeq
The integration in equation (\ref{eta}) is over a
region $U$ of the phase space of the colliding particles, which is 
specified by the energy-momentum conservation. 
It depends on the energy $p^{0}$ 
of the final state. 
Now we specialize to the case of a  process for which the reacting
particles are labeled by $l=1$ and $m=2$.
By using equation (\ref{spectral-F})
we can express $dF_{12}$ in terms of the spectral functions 
and the densities. This gives the following final expression
for the emission coefficient (i.e., the production rate) 
for electrons or photons:
\beq
\eta(\ee)=\frac{c\,n_{1}n_{2}}{16 \pi^{2}(1+\delta_{12})}\int_{U}\,
\:\prod_{j=1}^{2}\left[F_{j}(\ee_{j})d\ee_{j}\,d\Omega_{j}\:\right]\,
\fkin_{12}\,
\frac{d\sigma}{dP}.
\label{final-eta}
\eeq
\par Now we define the total reaction rate between two particles of
energies $\ee_{1}$ and $\ee_{2}$ to be
\beq
{\rm R}(\ee_{1},\ee_{2})= \frac{c\,n_{1} n_{2}}{2(1+\delta_{12})}
\int_{-1}^{1}d\mu\, \fkin_{12}\,\sigma_{\rm total},
\label{reacion-rate}
\eeq
where $\mu$ is the cosine of the angle between the momenta of the 
colliding particles and $\sigma_{\rm total}$ is the total cross section 
for the process considered (integrated over the entire phase space of the 
emitted particle). Clearly $\:\fkin_{12}$ as well as $\sigma_{\rm total}$
depend only on $\ee_{1}$, $\ee_{2}$, and $\mu$. Now it is possible to 
express the emission coefficient in terms of the total reaction rate as
\beq
\eta(\ee)=\int\,\prod_{j=1}^{2}[\,d\ee_{j}F_{j}(\ee_{j})\,]{\rm R}(\ee_{1},
\ee_{2}){\rm P}(\ee_{1},\ee_{2};\ee),
\label{coppi-method}
\eeq
where the integration is over all values of $\ee_{1}$ and $\ee_{2}$
without any restriction (in contrast with eq.[\ref{final-eta}]).
In the above equation, P is the probability, integrated over all
incident and emergent angles of the particles, for  emitting 
a particle of energy $\ee$, from a collision between the particles of
energies $\ee_{1}$ and $\ee_{2}$. It is normalized so that
$\int d\ee{\rm P}(\ee_{1},\ee_{2};\ee)=1$, where the integration 
is over all values of $\ee$. Equation (\ref{coppi-method}) has been
used by several previous authors (e.g., CB90). 
\par We can obtain the absorption coefficient from equation (\ref{eta})
with only  minor changes.
For the absorption of the particles of type $i$ with a momentum $p_{i}$
we find that
\beq
f_{i}(p_{i})\:\chi_{i}(p_{i})=\sum_{j}\frac{c}{1+\delta_{ij}}\int_{U}\,
\frac{dF_{ij}}{dP_{i}}\,\fkin_{ij}\,\sigma_{\rm total}.
\eeq
Here $\sigma_{\rm total}$ 
is the total scattering cross section for the process. The summation extends 
over all relevant processes.  For a binary process, 
involving the particles of type $i$ and type $j$, the   absorption 
coefficient can be written in terms of the spectral functions as follows:
\beq
\chi_{i}(\ee_{i})=\frac{c\,n_{j}}{4 \pi (1+\delta_{ij})}\,\int_{U}\,
d\ee_{j}d\Omega_{j}\,F_{j}(\ee_{j})\,\fkin_{ij}\,\sigma_{\rm total},
\label{final-chi}
\eeq
where the integration region $U$ is determined by the energy--momentum 
conservation. This way of writing the emission and absorption coefficients
is very convenient for Monte Carlo evaluation we describe below. 

We remark that in equations (\ref{final-eta}$-$\ref{final-chi})
we have used $\ee$ in a generic way and it has to be replaced by $\gamma$
whenever it refers to the pairs. Physically, $4\pi \ee^{2}\eta(\ee)$
is the rate at which photons of energy $\ee$ are
emitted per unit volume and unit energy due to the
process under consideration; similarly, $4\pi\,\beta\,\gamma^{2}\,\eta(\gamma)$
gives the corresponding electron emission rate (recall  that
we express energy in units of $mc^{2}$). Electron and photon 
absorption rates are obtained in a similar
way. If the size of the system is $l$, the optical depth $\tau$ and the 
absorption coefficient are related by $\tau=l \chi/c$. 
Equations (\ref{final-eta}) and (\ref{final-chi}) constitute the
point of departure for the following  two sections where we obtain the 
emission and  the absorption coefficients for the photon and the pair
kinetic equations. We remark here that in the case of Compton scattering
of the photons as well as the pairs, the collision integrals only give
the rate at which the spectrum changes at a given energy and do not imply
any change in the total numbers of the particles. 
\medskip
%
%
%
%
%
%
%
%
\mysection{3 Collision integrals for photons}
\setcounter{section}{3}
\reset
\smallskip

The preceding discussion has been very general. 
We now obtain the integral  expressions
for the photon emission coefficients due to Compton scattering, two-photon
pair annihilation, and bremsstrahlung and
the absorption coefficients due to Compton scattering
and the  pair creation. In this paper we do not consider the double-Compton 
emission
or the three-photon emission through pair annihilation . Also we 
do not consider the effect of 
photon absorption through the inverse-bremsstrahlung (free-free absorption).
\mysubsec{3.1 Compton scattering of photons}

The problem of Comptonization in astrophysics has 
been analyzed extensively by many previous authors (e.g., Blumenthal \& Gould
1970; Rybicki \& Lightman 1979; and more recently by CB90). 
Here we obtain an 
integral expression which is valid at all energies
of the incident electrons and photons. Throughout this paper we call the 
comoving frame of the plasma the C-frame. 
Let  $p$ and  $p_{1}$ be the momenta of the incident electron and
photon, respectively in the C-frame. Let $q$ and $q_{1}$ be
the corresponding momenta  after the scattering. Recall that
$p^{2}=1$ and $p_{1}^{2}=0$. We require the final photon energy to be
$\ee$. Hence we set $q_{1}=\ee(1,{\bf k})$, where $\bf k$ is the 
directional unit vector. We write  $p=\gamma(1,\bb)$
and $p_{1}=\ee_{1}(1,{\bf k}_{1})$. Here $\gamma$ is the Lorentz factor of 
the incident electron, $\bb$ is its three-velocity in units of $c$,
$\ee_{1}$ is the energy of the incident photon, and ${\bf k}_{1}$ is its
directional unit vector. Using the fact that $(p+p_{1}-q_{1})^{2}=
q^{2}=1$ we obtain the well known relation between the initial and the final
photon energies viz., $\ee=\tilde{\ee}(\ee_{1})$ or
$\ee_{1}=\tilde{\ee}_{1}(\ee)$, where
\beq
\tilde{\ee}=\frac{a_{1}\gamma\ee_{1}}{a\gamma+b\,\ee_{1}}
{\quad \rm and \quad} 
\tilde{\ee}_{1}=\frac{a\gamma\ee}{a_{1}\gamma-b\,\ee}.
\label{e relations}
\eeq
Here $a=1-\bb \cdot \bf k$, $a_{1}=1-\bb \cdot {\bf k}_{1}$, and
$b = 1- \cos\theta$, while $\cos\theta={\bf k}\cdot{\bf k}_{1}$ gives 
the cosine of the photon scattering angle in the C-frame. 
Let $\mu=\cos{\theta}$ and the cosine of the 
angle between $\bb$ and $\bf k$ is defined to be 
$\mu^{\prime}$. The angle between the
planes formed by the pairs of vectors
$({\bf k},{\bf k}_{1})$ and $({\bf k},\bb)$ is defined to
be $\phi$. It is easy to 
show (see the appendix for further details) using equation (\ref{final-eta}) 
that the Compton emissivity for photons is given by 
\beq
\eta(\ee)=\frac{c\,\nph(n_{-}+n_{+})\re^{2}}{8\,\pi\ee^{2}}
\,\int_{U}(d\gamma\,d\mu\,d\mu^{\prime}\,d\phi)F_{e}(\gamma)F_{\gamma}
(\tilde{\ee}_{1})\left (\frac{\Delta}{2\,\gamma^{2}a\,\xi}\right ),
\label{compton eta}
\eeq  
where $\Delta=\xi^{2}-\xi\,\sin^{2}\theta^{\prime}+1$ and  
$\xi=a_{1}\gamma/(a_{1}\gamma-b\,\ee)$, while $\theta^{\prime}$ is the 
photon scattering angle in the rest frame of the incident electron
and $\re$ is the classical radius of an electron. 
The region of integration $U$ is defined by $\gamma_{\rm min}\leq\gamma\leq
\gamma_{\rm max}$, $-1\leq\mu,\,\mu^{\prime}\,\leq 1$, and $0\leq\phi
\leq\,2\pi$ subject to the condition that $\ee_{1\,{\rm min}}\leq\tilde
{\ee}_{1}\leq\ee_{1\,{\rm max}}$. Here $\gamma_{\rm min}$ and 
$\gamma_{\rm max}$
are the limiting electron or positron energies in the plasma. Similarly
$\ee_{1\,{\rm min}}$ and $\ee_{1\,{\rm max}}$ are the limiting photon energies.

Now we obtain the corresponding ``absorption'' coefficient
(as stated before, this is not a 
real absorption; the photons are scattered into
a different energy bin). Let $p=\ee(1,{\bf k})$ and
$q=\gamma(1,\bb)$ be the initial momenta of the photon and the 
electron, respectively. Various symbols have the same 
meaning as above. The photon energy in the rest frame of the incident 
electron is given by $x=p\cdot q=\gamma\ee(1-\beta\,\mu)$,
where $\mu$ is the cosine of the angle between the vectors $\bb$ and $\bf k$.
Now $n_{j}=n_{+}+n_{-}$, $\delta_{ij}=0$, 
$F_{j}=F_{e}$, $d\Omega_{j}=2\,\pi d\mu$,
$\fkin_{ij}=(1-\beta\,\mu)$, $\ee_{i}=\ee$, and $\ee_{j}=\gamma$.
Substituting these expressions into equation (\ref{final-chi}), we obtain
\beq
\chi(\ee)=\frac{c\,(n_{-}+n_{+})}{2}\int_{U}\,d\mu\,d\gamma 
F_{e}(\gamma)(1-\beta\mu)\sigma_{\rm total}(x),
\label{compton-chi}
\eeq
where
\beq
\sigma_{\rm total}(x)=2\pi\re^{2}\left\{\frac{1+x}{x^{3}}\left[
\frac{2 x (1+x)}{1+2x}-\ln{(1+2x)}\right]
+\frac{\ln{(1+2x)}}{2x}-\frac{1+3x}{(1+2x)^{2}}\right\}
\label{compton-total}
\eeq
is the total cross section for Compton scattering (e.g., Jauch \& Rohrlich 
1980 -- JR80 henceforth). 
The integration domain $U$ is defined by $\gamma_{\rm min}\leq
\gamma\leq\gamma_{\rm max}$ and $ -1\leq\mu\leq 1$ 
without any restriction. Here
$\gamma_{\rm min}$ and $\gamma_{\rm max}$ 
are the limiting electron energies, as in the previous case.
\vspace{-10pt}
\mysubsec{3.2 emission and annihilation of photons by the pairs}

The emissivity due to the annihilation of relativistic electron-positron pairs
(creating two photons) has been analyzed by many authors
before (e.g., Zdziarski 1980; Ramaty \& M\'esz\'aros 1981; 
Yahel \& Brinkmann 1981; Svensson 1982a -- henceforth S82a).
We give here the final result using the notation of S82a and refer the
reader to that paper for a detailed derivation.
Let $p_{i}=\gamma_{i}(1,\bb_{i})$; $i=1,2$ be the momenta of the electron and
the positron, respectively in the C-frame. 
Let $q_{1}=\ee(1,{\bf k})$ be the momentum of
one of the emitted photons. Here $c\,\bb_{i}$ are the particle velocities
and $\gamma_{i}$ are the corresponding 
Lorentz factors, $\ee$ is the photon energy,
and $\bf k$ is its directional unit vector. The momentum of the C-frame 
itself is denoted by $q=(1,{\bf 0})$. We call the center-of-momentum 
frame of the pair the CM-frame and the quantities in this frame appear 
with a suffix `cm'. The particle momenta in this frame are
$p_{1\,{\rm cm}}=\gamma_{\rm cm}(1,\bb_{\rm cm}),
\,p_{2\,{\rm cm}}=\gamma_{\rm cm}(1,-\bb_{\rm cm}),\,
q_{1\,{\rm cm}}=\ee_{\rm cm}
(1,{\bf k}_{\rm cm})$, and $q_{\rm cm}=\gamma_{\rm c}(1,-\bb_{\rm c})$.
Here $\gamma_{\rm cm}$ is the Lorentz factor of the electron or positron, 
$\ee_{\rm cm}$ is the photon energy, and ${\bf k}_{\rm cm}$ is
its directional unit vector (in the CM-frame). The velocity of the CM-frame
as measured in the C-frame is $c\,\bb_{\rm c}$ and $\gamma_{\rm c}$ is 
the corresponding Lorentz factor. 
Various directional cosines are defined as follows:
$\mu,x,y$, and $z$ are the cosines of the angles between the pairs of
vectors ($\bb_{1}$, $\bb_{2}$),
(${\bf k}_{\rm cm}$, $\bb_{\rm cm}$), ($\bb_{\rm c}$, $\bb_{\rm cm}$),
and ($\bb_{\rm c}$, ${\bf k}_{\rm cm}$), respectively; the angle
between the planes formed by the pairs of  vectors
$(\bb_{\rm c},{\bf k}_{\rm cm})$ and
$(\bb_{\rm c},\bb_{\rm cm})$ is denoted by $\phi$. After analyzing the 
kinematics, we obtain
$\gamma_{\rm cm}=\surd{\left[
1/2+\gamma_{1}\gamma_{2}(1-\beta_{1}\beta_{2}\mu)/2\right]}$, 
$\gamma_{\rm c}=(\gamma_{1}+
\gamma_{2})/(2\,\gamma_{\rm cm})$, 
$y=(\gamma_{1}-\gamma_{2})/(2\,\beta_{\rm c}\,\beta_{\rm cm}\,\gamma_{\rm c}\,
\gamma_{\rm cm})$, 
$z=(\ee-\gamma_{\rm c}\gamma_{\rm cm})/(\beta_{\rm c}\gamma_{\rm c}
\gamma_{\rm cm})$, and $x=y z+\surd{[(1-y^{2})(1-z^{2})]} \cos{\phi}$.
Now using  equation (\ref{final-eta}) we obtain the pair 
emissivity
\beq
\eta(\ee)=\frac{c\,n_{+}n_{-}}{4\,\pi\,\ee^{2}}\,\int_{U}
d\mu\,d\phi\,\prod_{i=1}^{2}[F_{e}(\gamma_{i})\,d\gamma_{i}]
\,\frac{\beta_{\rm cm}\,\gamma_{\rm cm}}{\beta_{\rm c}\gamma_{\rm c}
\gamma_{1}\gamma_{2}}\,\left(\frac{d\sigma}{d\Omega}\right)_{\rm cm}.
\label{pair eta}
\eeq
The differential cross section in the CM-frame is given by
\beq
\left(\frac{d\sigma}{d\Omega}\right)_{\rm cm}=
\frac{\re^{2}}{4\,\beta_{\rm cm}\gamma^{2}_{\rm cm}}\,
\left[-1+\frac{3-\beta_{\rm cm}^{4}}{2}\left(\zeta_{+}+\zeta_{-}\right)-
\frac{1}{2\,\gamma^{4}_{\rm cm}}\left(\zeta_{+}^{2}+\zeta_{-}^{2}\right)
\right],
\label{pair diff cross}
\eeq
where $\zeta_{\pm}=1/(1\pm\beta_{\rm cm}x)$.
The integration domain
$U$ in equation (\ref{pair eta}) is given by $\gamma_{\rm min}\leq
\gamma_{1,2}\leq\gamma_{\rm max},\,-1\leq\mu\leq 1,\,\mbox{ and }0\leq\phi
\leq 2\pi$, subject to the condition $-1\leq z\leq 1$, which is 
equivalent to the condition $\Gamma_{-}(\gamma_{1},\gamma_{2},\mu;\ee)
\leq\gamma_{\rm cm}(\gamma_{1},\gamma_{2},\mu)
\leq\Gamma_{+}(\gamma_{1},\gamma_{2},\mu;\ee)$, where
$\Gamma_{\pm}=\ee\gamma_{\rm c}(1\pm\beta_{\rm c})$. Here
$\gamma_{\rm min}$ and $\gamma_{\rm max}$ are the limiting
pair energies in the plasma. 

Now we obtain the photon absorption coefficient due to  pair creation. 
Let the initial momenta of the photons be $p=\ee(1,{\bf k})$
and $p^{\prime}=\ee^{\prime}(1,{\bf k}^{\prime})$, with the usual meaning
for various symbols. If an electron-positron pair is produced,
then the  CM-frame Lorentz factor of the electron is given by $\gamma_{\rm cm}
=\surd{[\ee\ee^{\prime}(1-\mu)/2]}$, where 
$\mu$ is the cosine of the angle between the vectors
$\bf k$ and ${\bf k}^{\prime}$. Using equation (\ref{final-chi}) we find
\beq
\chi(\ee)=\frac{c\,n_{\gamma}}{4}\int_{U}\,d\mu\,d\ee^{\prime}
F_{\gamma}(\ee^{\prime})(1-\mu)\sigma_{\rm total}(\gamma_{\rm cm}).
\label{pair-chi}
\eeq
Since $\sigma(\gamma \gamma \rightarrow{\rm ee})=2\beta_{\rm cm}^{2}\sigma(
{\rm ee} \rightarrow \gamma \gamma)$, by integrating equation 
(\ref{pair diff cross}), we find 
\beq
\sigma_{\rm total}(\gamma_{\rm cm})=
\frac{\pi \re^{2} \beta_{\rm cm}}{\gamma_{\rm cm}^{2}}\left[\,
\frac{(3-\beta_{\rm cm}^{4})}{\beta_{\rm cm}}\,
{\rm ln}\left(\frac{1+\beta_{\rm cm}}{1-
\beta_{\rm cm}}\right)-2-\frac{2}{\gamma_{\rm cm}^{2}}\right].
\label{total-cross-sect-pair}
\eeq
The integration domain $U$ in equation (\ref{pair-chi}) 
is defined by $-1\leq\mu\leq 1$
and $\ee^{*}\leq\ee^{\prime}\leq\ee_{\rm max}$, where
$\ee^{*}= 2/[\ee(1-\mu)]$ is the pair creation threshold
energy and $\ee_{\rm max}$ is the limiting photon energy in 
the plasma. 
\mysubsec{3.3 bremsstrahlung emissivity}

The bremsstrahlung emissivity of a pair plasma has been analyzed 
in several papers (e.g., Haug 1975b, 1985c, 1987, 1989 and Dermer 1986).
The final expression for the photon emissivity can be written as
\beq
\eta_{\rm pair}(\ee)=\frac{c\,\alpha\re^{2}}{8\,\pi^{2}\,\ee}
\int_{U}\,d\mu\,d\Omega\,\prod_{i=1}^{2}[F_{e}(\gamma_{i})\,d\gamma_{i}]\:
\frac{\fkin_{12}}{\rho}\,
\left[\frac{1}{2}\left(n_{+}^{2}+n_{-}^{2}\right)\frac{C_{1}}
{\Delta_{1}}+n_{+}n_{-}\frac{C_{2}}{\Delta_{2}}\right],
\label{brems pair eta}
\eeq
where $\alpha$ is the fine structure constant. 
The first term inside the braces represents the sum of the electron-electron
and the positron-positron contributions and the second term gives the 
electron-positron contribution. The expressions for 
$\rho$, $\fkin_{12}$, $C_{i}$, and $\Delta_{i}$, 
along with the definitions of the integration variables $\mu$
and $\Omega$ are given in the appendix. The emissivity due to pair-proton
bremsstrahlung can be written as
\beq
\eta_{\rm proton}(\ee)=\frac{c\,
n_{\rm p}\left(n_{+}+n_{-}\right)}{4\,\pi\ee^{2}}
\int_{1+\ee}^{\gamma_{\rm max}}d\gamma\,F_{e}(\gamma)\beta\,
\left(\frac{d\sigma}{d\ee}\right)_{\rm proton},
\label{brems-proton}
\eeq
where $(d\sigma/d\ee)_{\rm proton}$ is the cross section for this process
(see e.g., JR80). Here the protons are assumed to be at rest. 
\begin{figure}[p]
\centerline{\psfig{figure=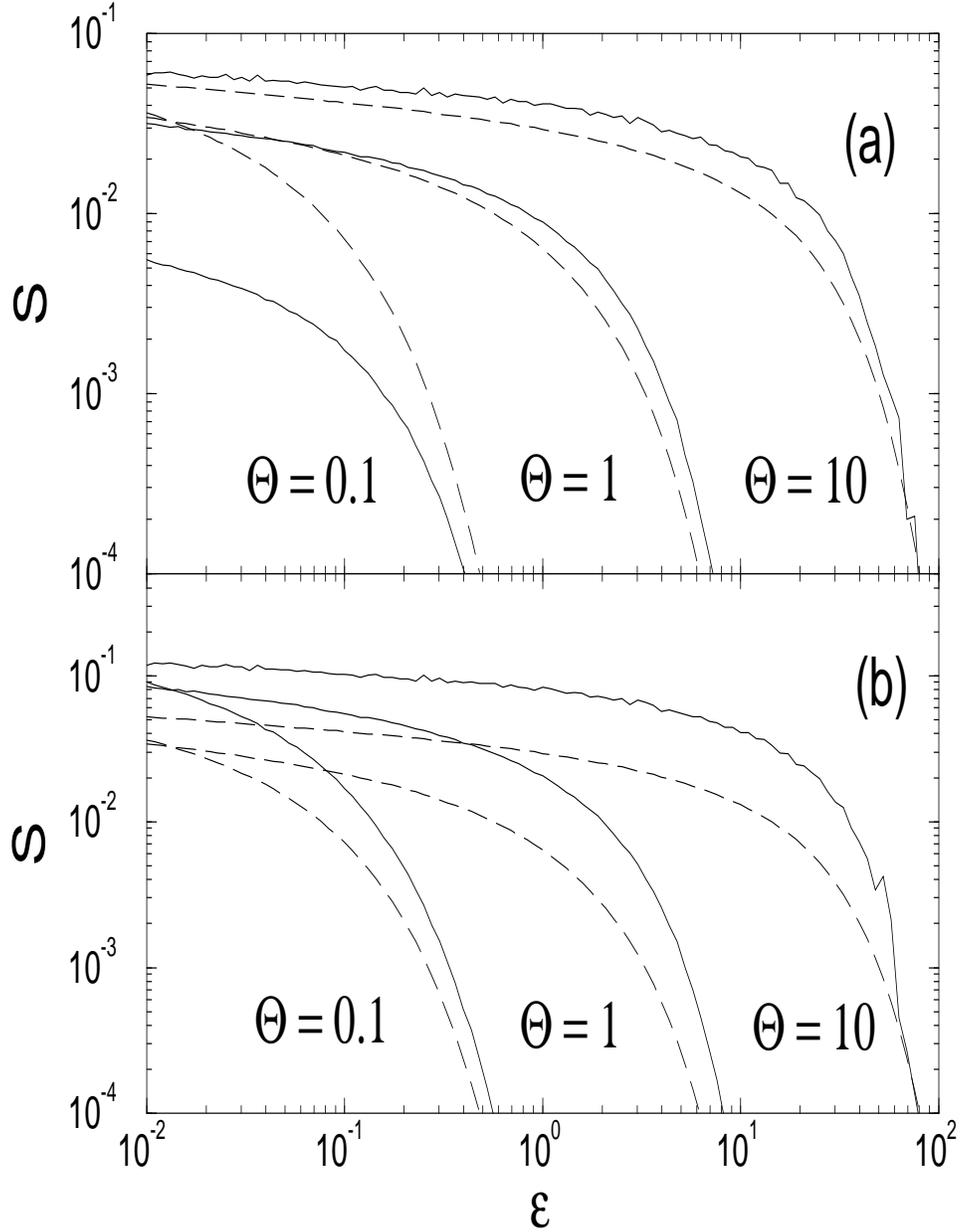,height=6.5in,width=5.5in}}
\caption{The emissivity due to (a) electron-electron and (b) electron-positron
bremsstrahlung from a thermal plasma for three different temperatures. 
The dashed lines represent the emissivity due to 
pair-proton bremsstrahlung (given here for comparison). The energy of the 
emitted photon is $\ee$ and $S = 4 \pi \ee^{3}\eta(\ee)/(c n_{1} n_{2} \sigma_{
\rm Th})$, where $n_{1,2}$ are the appropriate densities. These results
agree with Haug (1985c) and Dermer (1986).}
\end{figure}
\medskip
%
%
%
%
%
%
%
%
%
\mysection{4 Collision integrals for pairs}
\setcounter{section}{4}
\reset
\vspace{10pt}
\mysubsec{4.1 Compton scattering of the pairs}

The effect of Compton scattering on the pair distribution can be analyzed 
in a manner similar to that of the photon Comptonization 
discussed in the previous
section. Let $p$ and $p_{1}$ be the momenta of the incident 
electron and the 
photon, respectively. Let $q$ and $q_{1}$ be their corresponding 
momenta after the 
scattering. We require the final electron energy to be $\gamma$.
Hence, we set $q=\gamma (1,\bb)$, where $c \bb$ is the velocity of
the scattered electron. We write $p=\gamma_{1}(1,\bb_{1})$,
$p_{1}=\ee_{1}(1,{\bf k}_{1})$, and $q_{1}=\ee(1,{\bf k})$. 
Using $q_{1}^{2}= (p+p_{1}-q)^{2}=0$, we obtain a relation between the initial
energy of the photon and the final energy of the electron given by $\ee_{1}=
\tilde{\ee}_{1}$, where
\beq
\tilde{\ee}_{1}=\frac{b\,\gamma\,\gamma_{1}-1}{a_{1}\,\gamma_{1}-a\,\gamma}.
\label{e-rel}
\eeq
In this equation $a=1-\bb\cdot{\bf k}_{1}$, $a_{1}=1-\bb_{1}\cdot{\bf k}_{1}$,
and $b=1-\bb\cdot\bb_{1}$. Let $\mu$ be the cosine
of the angle between the vectors 
$\bb$ and ${\bf k}_{1}$. The cosine of the angle
between the vectors $\bb$ and $\bb_{1}$ is defined to be $\mu^{\prime}$.
The angle between the planes formed by the 
pairs of vectors $(\bb,{\bf k}_{1})$ and $(\bb,\bb_{1})$ is defined to be
$\phi$. Now the emission coefficient due to Compton scattering can be
written (see the appendix for details) as
\beq
\eta(\gamma)=c\,\nph\,(n_{-}+n_{+})\,\re^{2}\int
d\mu d\mu^{\prime}
 d\phi d\gamma_{1}\,F_{e}(\gamma_{1})\,F_{\gamma}(\tilde{\ee}_{1})
\,\frac{a_{1} X}{16\pi\,\ee\,\gamma\,\rho_{1}}
\left|\frac{d\tilde{\ee}_{1}/d\gamma}{1+d\ee/d\gamma}\right|,
\label{e-comp-emis}
\eeq
where
\beq
X=\frac{\rho_{1}}{\rho_{2}}+\frac{\rho_{2}}{\rho_{1}}+2\,
\left(\frac{1}{\rho_{1}}-\frac{1}{\rho_{2}}\right)+
\left(\frac{1}{\rho_{1}}-\frac{1}{\rho_{2}}\right)^{2},
\label{e-x}
\eeq
while $\rho_{1} = a_{1}\tilde{\ee}_{1}\gamma_{1}$ and $\rho_{2}=a\tilde{\ee}_
{1}\gamma$. The integration region is given by 
$-1\,\le\, \mu,\,\mu^{\prime}\,\le\,1$, $0\,\le\,\phi\,\le\,2\pi$, and 
$\gamma_{\rm min} \le\gamma_{1}\le\gamma_{\rm max}$ subject to the 
condition that  $\ee_{\rm min}\le\tilde{\ee}_{1}\le
\ee_{\rm max}$. 

Next we consider the absorption coefficient due to 
Compton scattering.  Let $p=\ee(1,{\bf k})$ and
$q=\gamma(1,\bb)$ be the initial momenta of the photon and the electron,
respectively. The photon energy in the rest frame of the incident electron
is given by $x=p\cdot q=\gamma \ee(1-\beta\mu)$, where $\mu$ is the
cosine of the angle between the vectors 
$\bb$ and $\bf k$. As in section $3.1$, it can be shown that
\beq
\chi(\gamma)=\frac{c\,\nph}{2}\int\,d\mu\,d\ee\,F_{\gamma}(\ee)\,
(1-\beta\mu)\,\sigma_{\rm total}(x),
\label{e-comp-abs}
\eeq
where $\sigma_{\rm total}$ is given by equation (\ref{compton-total}).
The integration domain is given by $-1\le \mu\le 1$ and 
$\ee_{\rm min}\le\ee\le\ee_{\rm max}$. 
\mysubsec{4.2 production and annihilation of the pairs}

The analysis
for this case is analogous to that for the 
pair annihilation emissivity discussed
above. Let $p_{i}=\ee_{i}(1,{\bf k}_{i})$ be the momenta
of the photons in the C-frame, where $\ee_{i}$ are their energies and
${\bf k}_{i}$ are their directional unit vectors. Let $p=\gamma(1,\bb)$ 
be the momentum of one of the emitted particles. Here $c\,\bb$ is its
velocity in the C-frame and $\gamma$ is the corresponding Lorentz factor. The
momentum of the C-frame itself is denoted by $q=(1,{\bf 0})$.
We denote the CM-frame quantities with a suffix `cm'. Let 
$p_{1 \rm cm}=\ee_{\rm cm}(1,{\bf k}_{\rm cm})$,
$p_{2 \rm cm}=\ee_{\rm cm}(1,-{\bf k}_{\rm cm})$,
$p_{\rm cm}=\gamma_{\rm cm}(1,\bb_{\rm cm})$,
and $q_{\rm cm}=\gamma_{\rm c}(1,\bb_{\rm c})$ represent $p_{1},p_{2},
p$, and $q$, respectively in the CM-frame. 
The velocity of the C-frame as measured in the CM-frame is $c \bb_{\rm c}$ 
and $\gamma_{\rm c}$ is the corresponding Lorentz factor. 
Various directional cosines are defined as follows: $\mu, x, y$, and $z$
are the cosines of the angles between the pairs of vectors $({\bf k}_{1},
{\bf k}_{2})$, $({\bf k}_{\rm cm},\bb_{\rm cm})$, $({\bf k}_{\rm cm},
\bb_{\rm c})$, and $(\bb_{\rm cm},\bb_{\rm c})$, respectively. The
angle between the planes formed by the pairs of vectors $(\bb_{\rm c},
\bb_{\rm cm})$ and $(\bb_{\rm c},{\bf k}_{\rm cm})$ is defined to be 
$\phi$.
We have $\gamma_{\rm cm}=\ee_{\rm cm}=\surd{[\ee_{1}\ee_{2}(1-\mu)/2]}$,
$\gamma_{\rm c}=(\ee_{1}+\ee_{2})/(2\,\ee_{\rm cm})$, $y=(\ee_{2}-\ee_{1})
/(2\,\beta_{\rm c}\,\gamma_{\rm c}\,\ee_{\rm cm})$, $z=(\gamma_{\rm c}
\gamma_{\rm cm}-\gamma)/\Delta$, whereas 
$\Delta=\beta_{\rm c}\beta_{\rm cm}\gamma_{\rm c}\gamma_{\rm cm}$, and
$x=y z+\surd{[(1-y^{2})(1-z^{2})]}\cos{\phi}$.
We can now write (see the appendix for more details) 
the pair creation emissivity as
\beq
\eta(\gamma)=\frac{c\,\nph^{2}}{16\pi\,\beta\gamma^{2}}
\int_{U}d\mu d\phi \prod_{i=1}^{2}
[F_{\gamma}(\ee_{i})d\ee_{i}]\:\frac{1-\mu}{\Delta}\:\left(
\frac{d\sigma}{d\Omega}\right)_{\rm cm},
\label{pair-cr-f}
\eeq
where the differential cross section is obtained by multiplying the one 
given by equation (\ref{pair diff cross}) with $\beta_{\rm cm}^{2}$.
The integration domain is given by $\ee_{\rm min}\leq \ee_{1,2}\leq
\ee_{\rm max}$, $-1\leq\mu\leq 1$, and $0\leq\phi\leq 2\pi$, subject
to the condition $-1\leq z\leq 1$, which is equivalent to 
$\Gamma_{-}\leq \gamma \leq \Gamma_{+}$, where $\Gamma_{\pm}=\gamma_{\rm c}
\gamma_{\rm cm}(1\pm \beta_{\rm c}\beta_{\rm cm})$.

For the absorption coefficient due to pair creation, consider an 
electron of momentum $p=\gamma(1,\bb)$
annihilating with a positron of momentum $p^{\prime}=\gamma^{\prime}(1,\bb^{
\prime})$. Their common Lorentz factor in the CM-frame is given by
 $\gamma_{\rm cm}=\surd{[\gamma\gamma^{\prime}
(1-\beta\beta^{\prime}\mu)/2]}$,
where $\mu$ is the cosine of the angle between the vectors $\bb$ and $\bb
^{\prime}$. Setting $\ee_{i}=\gamma$, $\ee_{j}=\gamma^{\prime}$,
$n_{j}=n_{\pm}$,
$\delta_{ij}=0$, $d\Omega_{j}=2\pi d\mu$, $F_{j}=F_{e}$, and $\fkin_{ij}=
\beta_{\rm r}\gamma_{\rm r}(\gamma \gamma^{\prime})^{-1}$ in
equation (\ref{final-chi}) we find
\beq
\chi_{\pm}(\gamma)=\frac{c\,n_{\mp}}{2}\int d\gamma^{\prime} d\mu F_{e}
(\gamma^{\prime})\:\frac{\beta_{\rm cm}\gamma_{\rm cm}^{2}}{\gamma\gamma^
{\prime}}\:\sigma_{\rm total}(\gamma_{\rm cm}).
\label{pair-abs-a}
\eeq
The integration is over the region $\gamma_{\rm min}\leq\gamma^{\prime}\leq
\gamma_{\rm max}$ and $-1\leq\mu\leq 1$ without any restriction. Here the 
limiting energies of the pairs are denoted by
$\gamma_{\rm min}$ and $\gamma_{\rm max}$.
Finally $\sigma_{\rm total}$ is the total
cross section for the pair annihilation, which is obtained by dividing the 
one given by equation (\ref{total-cross-sect-pair}) with 
$2 \beta_{\rm cm}^{2}$
\mysubsec{4.3 bremsstrahlung cooling rate}

Since
this process is much slower than all other reactions (roughly by a factor of 
$\alpha$ -- the fine structure constant) we can treat it to be continuous
in the energy and  momentum 
(i.e., $\frac{\Delta \gamma}{\gamma}\ll 1$) 
and use a continuity equation to describe it. 
At any time $t$, the density of electrons in the energy interval
$(\gamma,\gamma+d\gamma)$ is given by $\ne F_{e}(\gamma) d\gamma$. Clearly 
$\ne F_{e}(\gamma) \dot{\gamma}(\gamma)$ 
is the flux density of the electrons entering
this interval and $\ne F_{e}(\gamma+d\gamma)\dot{\gamma}(\gamma+d\gamma)$ is 
that due to the electrons leaving this interval (notice that
$\dot{\gamma}$ is negative in the case of electron cooling). 
The net contribution
to the electron or positron kinetic equation is now given by
\beq
\frac{\partial}{\partial t}\,[\ne(t)F_{e}(\gamma,t)]=-
\frac{\partial}{\partial\gamma}[\ne (t)F_{e}(\gamma,t)\dot{\gamma}]\equiv
C(\gamma,t).
\label{cool-a}
\eeq
The right hand side of this equation is essentially $4\pi\beta\gamma^{2}
(\eta-\chi f)$ for the process. The cooling rate $|\dot{\gamma}|$ can be 
written as the sum
\beq
|\dot{\gamma}|=E_{ep}(\gamma)+\int_{1}^{\infty}d\gamma^{\prime}F_{e}
(\gamma^{\prime})\left[
E_{ee}(\gamma,\gamma^{\prime})+E_{e\bar{e}}(\gamma,\gamma^{\prime})\right].
\label{cool-b}
\eeq
The cooling rates $E_{ee}, E_{e\bar{e}}$, and $E_{ep}$ for   
${\rm e}^{\pm}$-${\rm e}^{\pm}$,
${\rm e}^{\pm}$-${\rm e}^{\mp}$, and ${\rm e}^{\pm}$-proton processes,
respectively, 
are given in the  appendix.
\begin{figure}[p]
\centerline{\psfig{figure=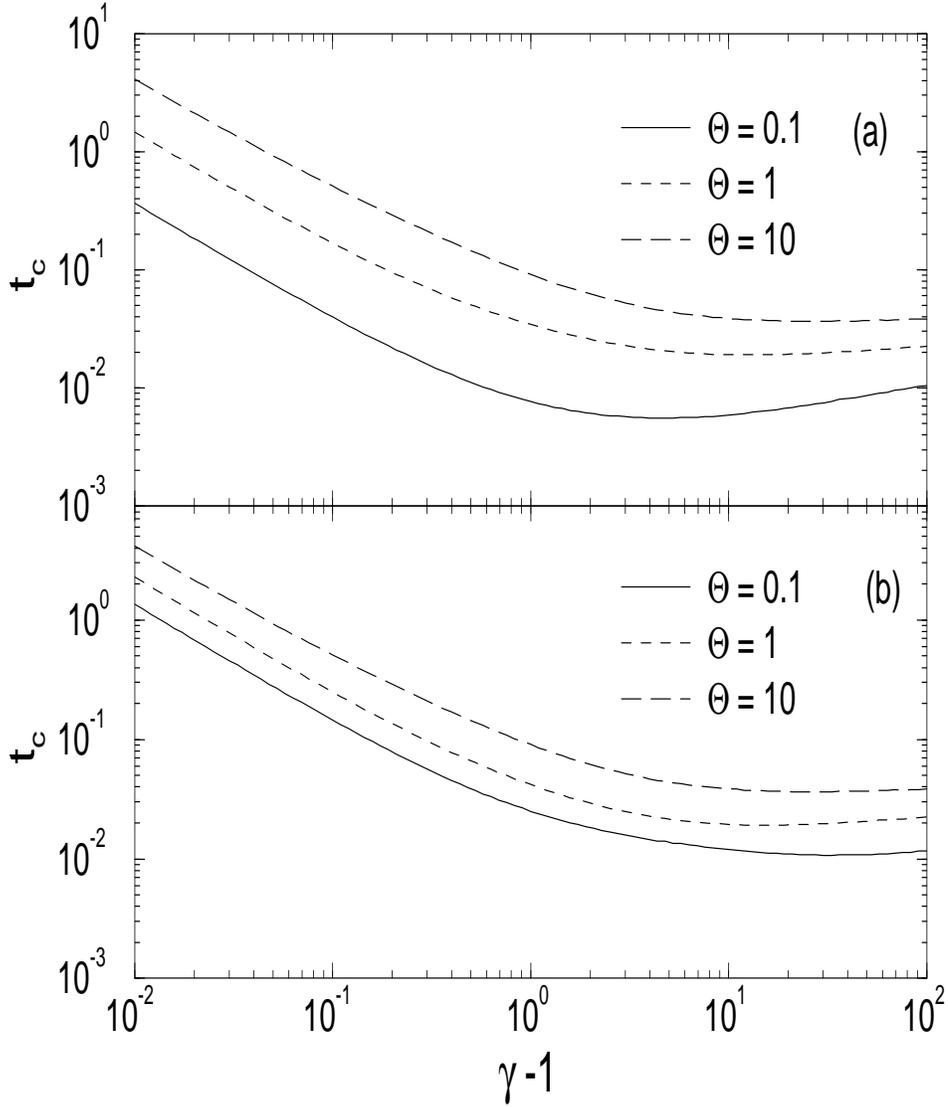,height=6.5in,width=5.5in}}
\caption{Bremsstrahlung cooling time for (a) an electron and (b) a positron
of energy $\gamma$ in a background thermal plasma (of electrons only)
of density $n_{\rm e}$. 
In the former case we use the cooling rate $E_{ee}$ and we use 
$E_{e\bar{e}}$ for the latter. Remaining cooling rates vanish in this
particular example.
The dimensionless cooling time is defined by
$t_{\rm c}=|\dot{\gamma}|/(cn_{\rm e}\sigma_{\rm Th}\gamma)$. Since the 
main time scale in the kinetics of the plasma is $\approx 
(cn_{e}\sigma_{\rm Th})$, $t_{\rm c}\ll 1$ at higher energies means that 
bremsstrahlung cooling is not very efficient at these energies.}
\end{figure}
\mysubsec{4.4 the effect of coulomb collisions}

Finally we analyze the effect of Bhabha and M\o ller collisions
(collectively termed as Coulomb collisions) on the electron 
spectrum (see e.g., Baring 1987b or CB90 for a similar treatment
and Dermer \& Liang 1989 for a Fokker--Planck treatment of this problem).
Here we ignore the diffusion term 
(which arises from the second order derivatives with respect to energy) 
that would
arise in the Fokker--Planck expansion of the kinetic equation as well as
the contribution from the pair-proton
collisions (which is a much slower process). Consider an
elastic scattering in which an electron with momentum $\bf p$ exchanges a 
momentum  $\bf q$ with a target particle in the plasma which is
either an electron or a positron. In both cases the collision cross
section diverges for $|{\bf q}|\rightarrow 0$ and it falls off rapidly 
for larger values of $|\bf q|$. We define $\theta$ to be the angle 
by which the incident electron is scattered. Small values of $|\bf q|$
correspond to the small angle collisions $(\theta\ll 1)$. More precisely,
$|{\bf q}|\cong |{\bf p}|\,\theta$ when $\theta$ is small. It is 
well known that the  divergence of 
the cross section for $\theta \rightarrow 0$ results in the domination of the
relaxation process by the scattering events with small angular deflections.
In many situations we can completely ignore the contribution from the
collisions which are producing large angle deflections. 
Let $l_{\pi/2}$ be the distance an electron has to travel
in order that its mean-square deflection is $\cong \,\pi/2$ and suppose
$L_{\pi/2}$ is the distance it has to travel so that it is 
deflected by an angle of $\pi/2$ in a single scattering, with a probability
close to unity. It can be shown that   
$L_{\pi/2}\,\cong\,16\,\overline{\gamma}^{2}/(45\pi\,\ne\,\re^{2})$ and 
$L_{\pi/2}/l_{\pi/2}\,\cong\, 2\ln{\Lambda_{\rm C}}$,
where $\overline{\gamma}$ is the mean electron momentum in the background
plasma. The latter ratio, in a nonrelativistic plasma, turns out to be $8
\ln{\Lambda_{\rm C}}$ but the expression for $\Lambda_{\rm C}$ is different
in that case.
The Coulomb logarithm for a relativistic plasma can be shown to be 
$\ln{\Lambda_{\rm C}}\cong 37+(3 \ln{\overline{\gamma}}- \ln{\ne})/2$.
In this equation $\ne$ refers to the number of electrons per 
cubic-centimeter. We consider
only those plasmas for which $\ln{\Lambda_{\rm C}}>$ a few, which
means that only 
small-momentum-transfer collisions are relevant.
In this limit Bhabha and M\o ller cross sections are 
equal. Therefore, we do not distinguish between 
electrons and positrons in the foregoing analysis.
Consider two distributions $f_{1}$ and $f_{2}$ of electrons. The
Boltzmann equation for $f_{1}$ can
be written as a continuity equation in the momentum space as 
\beq
\frac{\partial}{\partial t}f_{1}({\bf p})=-\frac{\partial}{\partial p^{i}}
S_{1}^{i}({\bf p}),
\label{land-a}
\eeq
where $S_{1}^{i}$ is the flux vector in the momentum space (see the 
appendix for its definition). Combining equation (\ref{boltzmann}) with 
the above  continuity equation we obtain
\beq
[\eta(\gamma)-\chi(\gamma)f(\gamma)]_{1}=
C_{11}(\gamma)+C_{12}(\gamma),
\label{land-i}
\eeq
where
\beq
C_{1s}(\gamma)=4\pi^{2}c\re^{2}
\,\ln{\Lambda_{\rm C}}\,\beta\,\frac{\partial}{\partial \gamma}\int\,
d\gamma^{\prime}\,\beta\beta^{\prime}\gamma^{\prime 2}Q(\gamma,
\gamma^{\prime}),
\label{land-j}
\eeq
while
\beq
Q(\gamma,\gamma^{\prime})=
\left[f_{1}(\gamma)\frac{\partial}{\partial 
\gamma^{\prime}}f_{s}(\gamma^{\prime})-f_{s}(\gamma^{\prime})
\frac{\partial}{\partial \gamma}f_{1}(\gamma)\right]\,
\int_{-1}^{1}d\mu B_{0}(\gamma,\gamma^{\prime},\mu).
\label{land-k}
\eeq
\begin{figure}[p]
\centerline{\psfig{figure=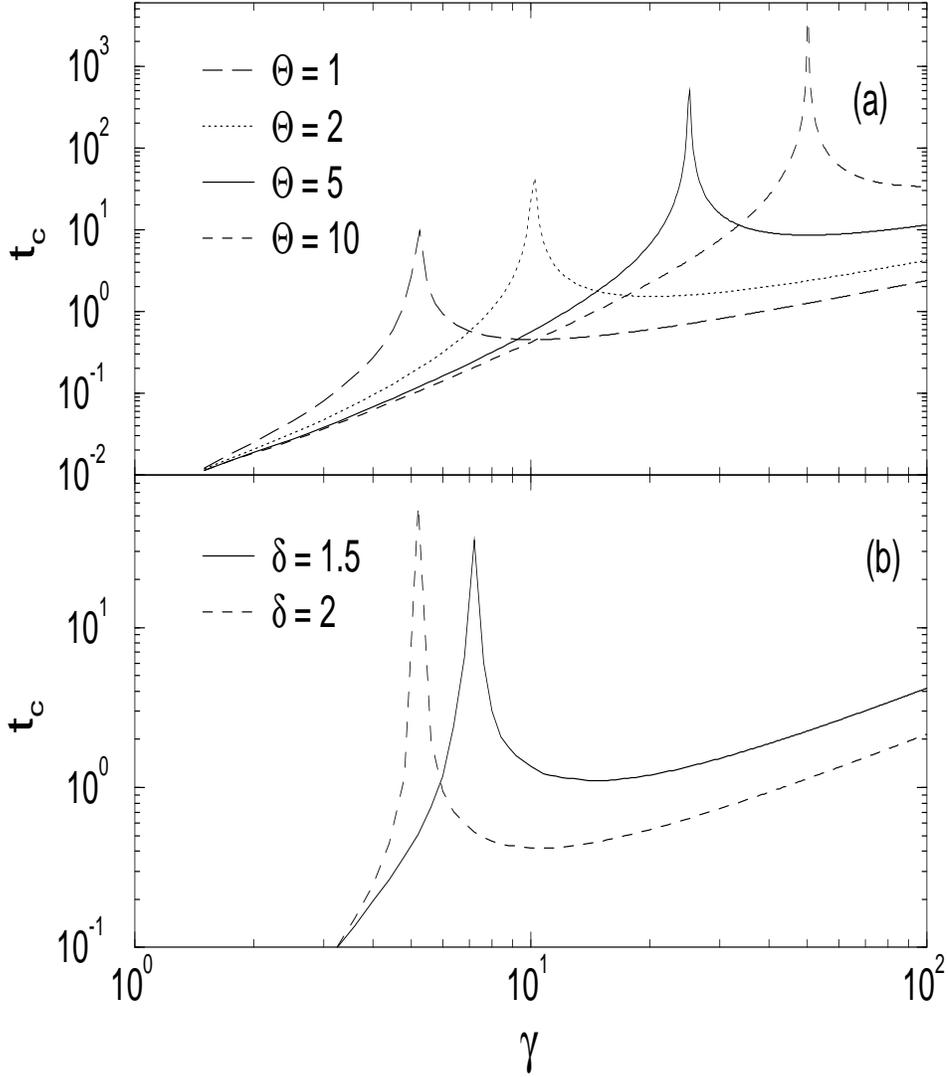,height=6.5in,width=5.5in}}
\caption{ 
Coulomb collision time for (a) a power law distribution with an index 
$\delta=2$ relaxing in a thermal background and (b) a power law distribution
relaxing through self-interactions. 
In both cases we have used $\ln{\Lambda_{\rm C}}=20$.
In general the time it takes to establish thermal equilibrium is many 
times that of the collision time. The spikes in these figures indicate that
the emission and absorption rates balance at that energy, because of the 
form of $C_{1s}$ (see the text).}
\end{figure}
The derivation of equation(\ref{land-j}), along with the definition of the 
quantities involved, is given in the appendix.
Here $C_{11}$ comes from the collisions within the electrons
of distribution $f_{1}$ and $C_{12}$ comes from the 
collisions of electrons of  distribution $f_{1}$ 
with the electrons of distribution $f_{2}$. 
In each case we have to use the appropriate electron density
in the Coulomb logarithm. Clearly, $C_{11}$ 
vanishes when $f_{1}$ is an equilibrium distribution. 
In Figure $3$ we give two examples of Coulomb relaxation. 
Recall that for electrons $f(\gamma)=n F(\gamma)/(4 \pi \beta \gamma^{2})$.
In the first example (Figure $3a$) we consider a nonthermal population of 
density $n_{1}$
and spectrum $F_{1}\propto \gamma^{-2}$ for $\gamma\ge 1$ interacting 
with a thermal background of density $n_{2}\gg n_{1}$ 
($f_{2}$ is the background distribution) and temperature $\Theta$.
The electrons relax mainly through collisions with the background and the 
reaction rate is determined by $C_{12}$ above ($C_{11}$ is negligible).
The dimensionless collision time is defined by $t_{\rm c} = c n_{2}\sigma_{\rm
Th} f_{1}(\gamma)/|C_{12}(\gamma)|$. In the second example (figure $3b$)
we consider a power law distribution of density $n_{1}$ and an index 
$\delta$, relaxing through 
self-interactions (there is no thermal background). In this case
$t_{\rm c} = c n_{1}\sigma_{\rm Th} f_{1}(\gamma)/|C_{11}(\gamma)|$.
\medskip
%
%
%
%
%
%
%
%
\mysection{5 The computational method}
\setcounter{section}{5}
\reset
\smallskip

In the kinetic theory approach to nonequilibrium plasmas that 
we have presented in
the preceding sections, the computational task is reduced to evaluating many
collision integrals (for each energy bin, after each time step) quickly and
efficiently, without compromising on the flexibility to handle many types of
distribution functions. Now we explain our new algorithm for adaptive Monte
Carlo integration which meets this demand. Our approach is similar to the
PSD method discussed in the introduction, with the principal difference being
that we are not using the conventional MC method (based on uniform sampling) 
to compute the transition probabilities (the collision integrals). There are
also some similarities between our method and the LP method described in the 
introduction. Both methods use statistical weights within a Monte Carlo 
scheme. In the LP method, these weights are introduced in an {\it ad hoc}
fashion, based on the energy carried by the LPs. In our method, we use
probability weights (see below) to enhance the sampling rate in those regions 
where the contribution to the integral being evaluated is greater. But these
weights (known as importance weights) are generated internally, through a 
minimal-variance prescription (see equations \ref{5.1.2d} and \ref{5.1.2f}).
Therefore, what we are using is a Monte Carlo method based on importance 
sampling. 

Being a kinetic-theory approach, our method bears a lot of resemblance to that
of C92. Both methods discretize the kinetic equations by following the particle
and photon statistics with reference to energy bins which are equally spaced
logarithmically. Both methods have similar simplifying assumptions about the
isotropy and the spatial uniformity of the photon and pair distribution 
functions. Our method is more flexible than that of C92 because of the adaptive
nature of the underlying Monte Carlo scheme (C92 uses
numerical integration to compute the collision kernels). However, there are 
several differences between these two methods.
It is straight forward to deal with 
anisotropic distributions in our method, except that some of the collision
integrals given in this paper require some modifications to incorporate them.
Moreover, we have not taken into 
account, the escape of photons and pairs from the system. We use the exact 
collision kernels throughout, as opposed to the ``two-moment'' approximations
used by C92 or other approximation schemes (see  CB90). In contrast to the 
method of C92, our time-evolution code is entirely dynamic i.e., we do not
precompute and store any quantities, thereby obviating the need to approximate
the final distribution functions after each time step. The integrands of some
of the collision kernels have very narrow peaks (``integrable singularities'')
which are hard to evaluate through numerical methods used before (e.g., C92
and references therein). Such ``singularities'' can be easily integrated by our
method. No special attention is required because the algorithm is adaptive --
it automatically adjusts the sampling rate, iteratively, to a high value at 
such points. Now we describe our computational method in detail.  
\mysubsec{5.1 discretization of the kinetic equations}

In order to simplify the analysis, we will consider only the case
for which $n_{+}=n_{-}=n_{\rm e}$ and $F_{+}(\gamma)=F_{-}(\gamma)=F_{\rm e}(
\gamma)$. This can be easily extended to the more general case. Let the 
net collision rate for the photons, due to Compton scattering, be given by
\beq
A_{\gamma}(\ee,t)=4\pi\ee^{2}\left[\eta_{\gamma}(\ee,t)-f_{\gamma}(\ee,t)
\chi_{\gamma}(\ee,t)\right]_{(e\gamma\leftrightarrow e\gamma)}.
\label{5.1.1a}
\eeq
The corresponding collision rate due to pair annihilation and creation
$(ee\leftrightarrow\gamma\gamma)$ is denoted by $B_{\gamma}(\ee,t)$. In an
analogous way we define $A_{\rm e}(\gamma,t)$ and $B_{\rm e}(\gamma,t)$
for the corresponding collision rates for 
pairs. From the photon rate equation
\beq
\frac{\partial}{\partial t}\left[n_{\gamma}(t)F_{\gamma}(\ee,t)\right]
=A_{\gamma}(\ee,t)+B_{\gamma}(\ee,t)
\label{5.1.1b}
\eeq
we obtain
\beq
\Delta F_{\gamma}(\ee,t)=
\frac{\left[A_{\gamma}(\ee,t)+B_{\gamma}(\ee,t)\right]\Delta t-F_{\gamma}
(\ee,t)\Delta n_{\gamma}(t)}{n_{\gamma}(t)+\Delta n_{\gamma}(t)},
\label{5.1.1c}
\eeq
where $\Delta F_{\gamma}(\ee,t)=F_{\gamma}(\ee,t+\Delta t)-F_{\gamma}(\ee,t)$,
$\Delta n_{\gamma}(t)= \dot{n}_{\gamma}(t)\Delta t$, and 
$\dot{n}_{\gamma}(t)$ is given below. Similarly we can obtain $\Delta F_{\rm e}(\gamma,t)$.
The time increment $\Delta t$ for each time step is chosen
in such a way that 
\beq
\left |\frac{\partial}{\partial t} f_{\gamma}(\ee,t)\right |\Delta t
\leq\nu f_{\gamma}(\ee,t)\qquad {\rm and}\qquad\left |
\frac{\partial}{\partial t}f_{\rm e}(\gamma ,t)\right |\Delta t
\leq f_{\rm e}(\gamma ,t)
\label{5.1.1d}
\eeq
for all values of $\ee$ and $\gamma$. 
In our computation we have used $\nu = 0.1$,
which means that the maximum change in $F_{\rm e}$ or $F_{\gamma}$  
in any energy bin, during any time step, is
less than or equal to $10\%$. Now we determine $\dot{n}_
{\gamma}$ arising from the pair processes (there is no change
in $\ne$ or $\nph$ arising from Compton scattering). We have
\beq
\int_{0}^{\infty}d\ee A_{\gamma}(\ee,t)=0\qquad{\rm and}\qquad\dot{n}_{\gamma}
(t)=\int_{0}^{\infty}d\ee B_{\gamma}(\ee,t)\label{5.1.1e}
\eeq
and two analogous equations for pairs. It can be shown that the positron
annihilation and creation rates are given by
\beq
\dot{n}_{\rm ann}(t)=\frac{c\ne^{2}(t)}{2}\int d\mu d\gamma d\gamma^{\prime}
F_{\rm e}(\gamma,t)F_{\rm e}(\gamma^{\prime},t)\frac{\beta_{\rm cm}\gamma^{2}_
{\rm cm}}{\gamma \gamma^{\prime}}\sigma_{ee\rightarrow\gamma\gamma}
\label{5.1.1f}
\eeq
and
\beq
\dot{n}_{\rm cr}(t)=\frac{c\nph^{2}(t)}{8}\int d\mu d\ee d\ee^{\prime}
F_{\gamma}(\ee,t)F_{\gamma}(\ee^{\prime},t)(1-\mu)
\sigma_{\gamma\gamma\rightarrow ee},
\label{5.1.1g}
\eeq
respectively. Here $\sigma_{ee\rightarrow\gamma\gamma}$ and 
$\sigma_{\gamma\gamma\rightarrow ee}$ are the total cross sections,
which are given in the previous sections. 
For each positron annihilated or created, there will be
a creation or annihilation, respectively, of two photons. Therefore we have
\beq
\dot{n}_{\gamma}(t)=2\left[\dot{n}_{\rm ann}(t)-\dot{n}_{\rm cr}(t)\right]
\qquad{\rm and}\qquad\dot{n}_{\rm e}(t) = \dot{n}_{\rm cr}(t)-\dot{n}_{\rm ann}
(t). \label{5.1.1i}
\eeq
We remark that $\dot{n}_{-}=\dot{n}_{+}=\dot{n}_{\rm e}$. We have verified
equations (\ref{5.1.1e}) and (\ref{5.1.1i}) 
in all our computations for time evolution, thereby ensuring
the number conservation.  In addition, we have verified the conservation 
of the total energy after each time step. 
Now we can use equation (\ref{5.1.1c})
iteratively, to obtain the time evolution of $F_{\rm e}$ and $F_{\gamma}$
from the initial data viz., $F_{\rm e}(\gamma,0)$, $F_{\gamma}(\ee,0)$,
and the initial densities. We have discretized the energy ($\ee$ and $\gamma$)
with twenty energy bins per decade of energy and used a logarithmic
interpolation
between these points to reconstruct $F_{\rm e}$ and $F_{\gamma}$ for 
the subsequent time steps, which are then used in the collision integrals. Now
the problem reduces to an  efficient evaluation of these multidimensional 
collision integrals with complicated integrands. For this purpose we have 
developed a new version of an adaptive and iterative Monte Carlo method. It
progressively adjusts itself to the nature of the integrand. 
We describe our algorithm below. 
\mysubsec{5.2 the adaptive monte carlo method}

A general purpose algorithm for multidimensional integration which
is widely used in the experimental particle physics is given by Lepage (1978).
It is an iterative and adaptive scheme. A computer program implementing
this method, known as VEGAS, can be found in Press et al. (1992). However,
we have found that it has several shortcomings when applied to the type 
of integrals that arise in the kinetic theory. Not only is the convergence 
weak in some cases, we have 
found that the subroutine gave erroneous output for the
high energy tails of the distributions. This is a significant obstacle 
because of the integrals over energy that we have to perform at the end of
each time step. That integration makes the errors propagate to lower
energies (where the results are otherwise accurate) during the  succeeding 
time steps. We will briefly explain the original method by Lepage and
then describe our modified scheme which can handle the integrals we need.
Firstly, by scaling the integration variable, any multidimensional 
integral can be written in the form 
\beq
{\cal I}=\int d{\bf r}f({\bf r}), \label{5.1.2a}
\eeq
where ${\bf r}=(z_{1},z_{2}, ..., z_{n})$, $d{\bf r}=\prod_{i=1}^{n}dz_{i}$,
and the integration is over the n-dimensional hypercube $0\leq z_{i}<1,
i=1,2,...,n$. If we generate $M\gg 1$ random points ${\bf r}_{k}$ with
a normalized probability density $p({\bf r})$ then the integral can be
approximated by
\beq
{\cal I}\simeq\frac{1}{M}\sum_{k}\frac{f({\bf r}_{k})}{p({\bf r}_{k})}.
\label{5.1.2b}
\eeq
The variance is given by
\begin{eqnarray}
\sigma^{2}[p]&=&\frac{1}{M-1}\left[\int d{\bf r}\frac{f^{2}({\bf r})}
{p({\bf r})} - {\cal I}^{2}\right]\nonumber\\
&\rightarrow&\frac{1}{M-1}\left[\sum_{k}
\frac{f^{2}({\bf r}_{k})}{p^{2}({\bf r}_{k})}-{\cal I}^{2}\right].
\label{5.1.2c}
\end{eqnarray}
The optimal choice for $p({\bf r})$ which 
minimizes the variance is derived from
\beq
\frac{\delta}{\delta p}\left\{\sigma^{2}[p]+\lambda\int d{\bf r}\,
p({\bf r})\right\}=0,\label{5.1.2d}
\eeq
which implies that 
\beq
p({\bf r}) = \frac{\left|f({\bf r})\right|}{\cal I}.\label{5.1.2e}
\eeq
This is the central theme of the importance sampling technique 
-- sample more in the
regions where the absolute value of the function is larger. However,
observe that the denominator is the integral itself! Thus we need an
algorithm to solve it  iteratively, starting with a reasonable 
guess for $p$. Then we calculate the integral 
by using equation (\ref{5.1.2b}) which then determines the 
new form for $p({\bf r})$, and so on.
If this process converges in a  manageable number of iterations, 
then we will have 
achieved our goal. The data storage requirements of directly implementing this
scheme are well within the reach of many present-day computers. The method by
Lepage consists of a restrictive assumption that the probability
density is separable. For instance, when $n=2$ and ${\bf r}=(x,y)$, the 
separability means that $p(x,y)=p_{x}(x)p_{y}(y)$ and to minimize the
variance we need
\beq
\frac{\delta}{\delta p_{x}}\left\{\sigma^{2}[p_{x},p_{y}]+
\lambda_{x}\int_{0}^{1} dx\,p_{x}(x)+\lambda_{y}\int_{0}^{1}dy\,p_{y}(y)
\right\}=0,\label{5.1.2f}
\eeq
which implies that 
\beq
p_{x}(x)=\frac{\left[\int_{0}^{1}dy\frac{f^{2}(x,y)}{p_{y}(y)}\right]^{1/2}}
{\int_{0}^{1}dx\left[\int_{0}^{1}dy\frac{f^{2}(x,y)}{p_{y}(y)}\right]^{1/2}},
\label{5.1.2g}
\eeq
and a similar equation for $p_{y}(y)$. For arbitrary dimensions, this scheme
is implemented in the VEGAS subroutine, mentioned before. The motivation for 
assuming the separability, according to Lepage(1978), is that it limits the
storage requirements. It is  not a good assumption in general. 
Therefore we proceed to implement importance sampling directly. All
essential features of the algorithm can be captured in a one dimensional
example which we will consider first. Then we will show how it can be
generalized to higher dimensions. Consider the integral ${\cal I}=\int_{0}^{1}
dx\,f(x)$. Let $p(x)$ be the normalized probability density we want. Suppose
$N$ is an integer greater than unity 
and $0=x_{0}<x_{1}<x_{2}<...<x_{N}=1$, while $\Delta x_{i}=
x_{i}-x_{i-1}$ for $i=1,2,...,N$. We will use the following discrete
representation of the probability density:
\beq
p(x)=\frac{1}{N\Delta x_{i}},\qquad{\rm if}\quad x_{i-1}\leq x<x_{i},
\label{5.1.2h}
\eeq
so that $\int_{x_{i-1}}^{x_{i}}\,dx\,p(x)=1/N$ for all $i$. Here the 
bin sizes $\Delta x_{i}$ need not be all equal but all bins have the same
probability weight. If the bin sizes are equal, we will get a uniform 
probability distribution leading to the crude Monte Carlo method. Now the
integral is approximated by ${\cal I}=\sum_{k=1}^{M}f(a_{k})/M
p(a_{k})$, where $0\leq a_{k} <1$ are uniformly distributed random numbers.
Typically $M\gg N$. Let
\beq
u_{i}=\frac{N}{M}\sum_{k=1}^{M}c_{i}(k)\left|f(a_{k})\right|,\label{5.1.2i}
\eeq
where $c_{i}(k)=1$, if $x_{i-1}\leq a_{k} < x_{i}$, and is zero otherwise.
Clearly, $\sum_{i=1}^{N}u_{i}\Delta x_{i} = {\cal I}$. Therefore $w_{i} = 
u_{i}\Delta x_{i}/{\cal I}$ is the importance weight associated with the
$i^{\rm th}$ bin. Since different bins contribute different amounts to
the integral, the idea now is to find a new set of bin spacings $\{x_{1},
x_{2}, ..., x_{N-1}\}$, so that all bins have equal importance weight
$w_{0}= 1/N$. Let $l$ be an integer (which depends on the bin location $i$)
such that 
\beq
\sum_{m=1}^{l}w_{m}\,\leq\,iw_{0}\,<\,\sum_{m=1}^{l+1}w_{m}.\label{5.1.2j}
\eeq
Then the new grid position for the $i^{\rm th}$ bin can be obtained from
\beq
x_{i,{\rm new}}=x_{l,{\rm old}}+\frac{1}{w_{0}}\left(iw_{0}-
\sum_{m=1}^{l}w_{m}\right)\left(x_{l+1,{\rm old}}-x_{l,{\rm old}}\right).
\label{5.1.2k}
\eeq
However, in practice we must damp the convergence so that the
contribution from the low-importance bins is not overly suppressed. As in the 
method by Lepage, we will damp the convergence by using the modified importance
weights given by
\beq
w^{\prime}_{i}=\left[\frac{1-w_{i}}{{\rm log}(1/w_{i})}\right]^{\alpha},
\label{5.1.2l}
\eeq
which gives $w_{0}^{\prime}=\sum_{i=1}^{N}w_{i}^{\prime}/N$. 
We now replace $w_{0}$ and $w_{i}$ with the corresponding primed 
quantities in the above equations. The new probability density is now
determined by using equation (\ref{5.1.2h}) and the process is repeated 
iteratively. If it converges, we will have $x_{i,{\rm new}}\cong
x_{i,{\rm old}}$ for all $i$, from which we can obtain the desired estimate
for $\cal I$. Now we give the extension of this scheme 
to two dimensions. We will
assume that the number of bins is $N$ for each dimension. A discrete 
representation of the probability density is given by
\beq
p(x,y)=\frac{1}{N^{2}\Delta x_{i}\,\Delta y_{j}},\qquad{\rm if}\quad
 x_{i-1}\leq x < x_{i}\quad{\rm and}\quad y_{j-1}\leq y < y_{j}.
\label{5.1.2m}
\eeq
This does not mean that the probability density is separable because $\Delta 
x_{i}$ and $\Delta y_{j}$ are not independent in general.
The integral is now estimated by
\beq
{\cal I}\simeq \frac{1}{M}\sum_{k=1}^{M}\frac{f(a_{k},b_{k})}{p(a_{k},b_{k})},
\label{5.1.2n}
\eeq
where $0\leq a_{k} <1$ and $0\leq b_{k} <1$ are uniformly distributed 
random numbers. 
Let
\beq
h_{ij}=\frac{N^{2}}{M}\sum_{k=1}^{M}c_{ij}(k)\left|f(a_{k},b_{k})\right|,
\label{5.1.2o}
\eeq
where $c_{ij}(k)=1$, if $x_{i-1}\leq a_{k} < x_{i}$ and $y_{j-1}\leq b_{k} <
y_{j}$, and is zero otherwise. Now we define $u_{i} = \sum_{j=1}^{N}h_{ij}
\Delta y_{j}$ and $v_{j}=\sum_{i=1}^{N}h_{ij} \Delta x_{i}$. Clearly,
${\cal I} = \sum_{i=1}^{N}u_{i} \Delta x_{i} = \sum_{j=1}^{N}v_{j}\Delta y_{j}
$. Let $w_{x_{i}}=u_{i}\Delta x_{i}/{\cal I}$ and $w_{y_{j}}=v_{j}
\Delta y_{j}/{\cal I}$. From these importance weights for $x$ and $y$ grids
we can obtain the corresponding damped weights and proceed to iterate as
if these were two one-dimensional problems. Generalization to arbitrary
dimensions is now trivial. 

           In all our applications we found that the values $N=70$ and
$\alpha = 1.3$ (for the damping index) gave stable and satisfactory results
within at most ten iterations or so. In general it is advisable to start
with a few thousand samples and after several iterations, increase $M$ 
(and retaining the resulting grid) and further iterate, and so on. For
many types of integrals, of at most five dimensions, we found that 
$M_{\rm max}\simeq 10^{4}$ samples to be adequate. For all the results
presented in this paper,
we have used the subroutine ran2 in Press et al.(1992) for the random
number generation. We find that our method is faster than the crude
Monte Carlo method (using uniform sampling) by a factor of ten or
better, which is also the case with the method by Lepage
(when it is applicable).
\medskip
%
%
%
%
%
%
%
%
\mysection{6 Time evolution and equilibria}
\setcounter{section}{6}
\reset
\smallskip
Here we give an analytical description of the equilibrium states of a pair
plasma, in terms of the initial conditions. For two specific examples, we 
follow the relaxation toward equilibrium using our time-evolution code. These
examples are meant to demonstrate that the whole formalism of this paper
(the collision integrals and the computational method) actually works.
We are considering a homogeneous, stationary, isotropic, and nonmagnetic 
system. There are no radiative transfer or hydrodynamic effects. On 
short time-scales $t\approx t_{\rm Th}=(n_{+}\sigma_{\rm Th}c)^{-1}$ the 
kinetics is determined by the rate-equations alone (see eq.[\ref{boltzmann}]).
We have seen that the collision integrals for these equations are nonlinear
functionals of the distribution functions. Given the initial state of the 
plasma, we can solve these first order coupled and nonlinear 
integro-differential 
equations to determine the time evolution of the distributions. The system
is characterized by the densities $n_{\gamma}$, $n_{+}$, and $n_{\rm p}$
and the spectra $F_{\gamma}(\ee)$ and $F_{\rm e}(\gamma)$, all of which
depend on time. Their values at $t=0$ define the initial state of the 
system. The total density of the particles is given by $\tilde{n}=n_{\gamma}+
2n_{+}+n_{\rm p}$ and the total energy density (including the rest energy of 
the pairs) is given  by $\tilde{u}= u_{\gamma}+u_{-}+u_{+}$, where
$u_{\gamma} = n_{\gamma}\int_{0}^{\infty}\ee F_{\gamma}(\ee)d\ee$ and 
$u_{\pm}=n_{\pm}\int_{1}^{\infty}\gamma F_{\rm e}(\gamma)d\gamma$. The mean
energy per particle is given by $\overline{\ee}=\tilde{u}/\tilde{n}$.
We see that there will be no change in $\tilde{n}$ due to Compton scattering 
or the pair annihilation and  creation. It will change only as a result of 
bremsstrahlung (also double Compton scattering and the pair annihilation into
three photons) which operate on a longer time scale $t\approx t_{\rm Th}/
\alpha$ ($\alpha$ is the fine structure constant). However $\tilde{u}$
remains constant throughout. Therefore we can divide the approach of the 
system toward equilibrium into two phases: $(i)$ The faster phase in which
both $\tilde{u}$ and $\tilde{n}$ remain constant and the system approaches to 
a state of kinetic equilibrium so that the total reaction rates for Compton 
scattering and the pair annihilation vanish (separately). This state is 
characterized by a temperature $\tilde{\Theta}$ and the chemical potentials
$\tilde{\mu}_{\gamma}$ and $\tilde{\mu}_{\pm}$  $(ii)$ The slower phase
in which $\tilde{u}$ is constant but $\tilde{n}$ changes, mainly due to 
bremsstrahlung (or its inverse, and other radiative processes)
so that the system finally reaches a
thermal equilibrium state characterized by a temperature $\Theta_{0}$
and a total density $n_{0}$. In this state the chemical potentials vanish
(see below). If $\Theta_{0}<\tilde{\Theta}$ then $n_{0}>\tilde{n}$, 
which means that this
phase is mainly the cooling of the plasma through bremsstrahlung and other
similar processes. On the 
other hand, if $\Theta_{0}>\tilde{\Theta}$ then the plasma will heat up due
to the inverse bremsstrahlung (free-free absorption) and other radiative
processes. 
\mysubsec{6.1 kinetic equilibrium: the densities and the temperature}

Consider Compton scattering of an electron of energy $\gamma$ and a photon
of energy $\ee$. The respective energies after the scattering are 
taken to be $\gamma^
{\prime}$ and $\ee^{\prime}$. If the total reaction rate vanishes, then
we have 
\beq
f_{\rm}(\gamma)f_{\gamma}(\ee)\left[1+\frac{\lambda_{0}^{3}}{2}f_{\gamma}
(\ee^{\prime})\right]= f_{\rm}(\gamma^{\prime})f_{\gamma}(\ee^{\prime})\left
[1+\frac{\lambda_{0}^{3}}{2}f_{\gamma}(\ee)\right], \label{5.3.3}
\eeq
where we have retained the Bose--Einstein enhancement factor for the photons
and $\lambda_{0}=h/mc$. The factor half in this equation takes
into account the polarization degeneracy of the photon states. Using the 
general form of the distribution functions 
\beq
f_{\gamma}(\ee)=\frac{2}{\lambda_{0}^{3}\left[{\rm exp}\left(\frac
{\ee - \mu_{\gamma}}{\Theta_{\gamma}}\right) - 1 \right]}\quad{\rm and}\quad
f_{\pm}(\gamma)=\frac{2}{\lambda_{0}^{3}}{\rm exp}\left(\frac{\mu_{\pm} - 
\gamma}{\Theta_{\pm}}\right)
\label{5.3.4}
\eeq 
and equation (\ref{5.3.3}) we find $\Theta_{+}=\Theta_{\gamma}=\Theta_{-}$.
We  denote this common temperature by $\widetilde{\Theta}$. Notice that 
equation (\ref{5.3.3}) does not yield any condition on the chemical potentials.
Now requiring that the total reaction rate should vanish for 
the pair annihilation and  creation as well, we find
\beq
f_{+}(\gamma_{+})f_{-}(\gamma_{-})
\left[1+\frac{\lambda_{0}^{3}}{2}f_{\gamma}(\ee_{1})\right]
\left[1+\frac{\lambda_{0}^{3}}{2}f_{\gamma}(\ee_{2})\right] = 
f_{\gamma}(\ee_{1})f_{\gamma}(\ee_{2}), \label{5.3.5}
\eeq
where $\gamma_{\pm}$ are the pair energies and $\ee_{1,2}$ are the photon
energies. Using the fact that the pairs and the photons have a common
temperature $\widetilde{\Theta}$, we obtain from this equation $\widetilde
{\mu}_{-}+\widetilde{\mu}_{+}= 2\widetilde{\mu}_{\gamma}$.
If there are no ions in the plasma 
(i.e.,  $n_{\rm p}= 0$) then $\widetilde{n}_{-} = \widetilde{n}_{+}$ so that 
$\widetilde{\mu}_{-}=\widetilde{\mu}_{+}=\widetilde{\mu}_{\gamma}$.
By assuming that exp$[(\ee - \tilde{\mu}_{\gamma})/\widetilde{\Theta}]
\gg 1$ and exp$[(\gamma - \tilde{\mu}_{\pm})/\widetilde{\Theta}]\gg 1$, 
for the relevant energies, we obtain the distribution functions
in the kinetic equilibrium state to be 
\beq
f_{\gamma}(\ee)=\frac{2}{\lambda_{0}^{3}}{\rm exp}\left(\frac{\tilde{\mu}_{
\gamma} - \ee }{\widetilde{\Theta}}\right)\quad{\rm and}\quad
f_{\pm}(\gamma)=\frac{2}{\lambda_{0}^{3}}{\rm exp}\left(\frac{\tilde{\mu}_{
\pm} - \gamma }{\widetilde{\Theta}}\right).
\label{5.3.6a}
\eeq
The densities are given by 
\beq
\tilde{n}_{\gamma}=\int_{0}^{\infty}4\pi \ee^{2} f_{\gamma}(\ee)d\ee = 
16 \pi \left(\frac{\widetilde{\Theta}}{\lambda_{0}}\right)^{3}{\rm exp}
\left(\frac{\tilde{\mu}_{\gamma}}{\widetilde{\Theta}}\right) \label{5.3.6}
\eeq
and
\beq
\tilde{n}_{\pm} = \int_{1}^{\infty}4 \pi \gamma \sqrt{\gamma^{2}-1} f_{\pm}
(\gamma)d\gamma = \frac{8\pi}{\lambda_{0}^{3}}\widetilde{\Theta}K_{2}\left(
\frac{1}{\widetilde{\Theta}}\right){\rm exp}\left(\frac{\tilde{\mu}_{\pm}}
{\widetilde{\Theta}}\right), \label{5.3.7}
\eeq
where $K_{n}$ is the $n^{\rm th}$ order modified Bessel function of the 
second kind. Using the relation $2\tilde{\mu}_{\gamma}=\tilde{\mu}_{-}+
\tilde{\mu}_{+}$ we find
\beq
\tilde{n}_{\gamma}=4\zeta^{2}\tilde{n}_{+}(n_{\rm p}+\tilde{n}_{+}), 
\label{5.3.8}
\eeq
where $\zeta = \widetilde{\Theta}^{2}/K_{2}(1/\widetilde{\Theta})$.
Finally, from the equation  $\tilde{n}_{\gamma}+2 
\tilde{n}_{+}+n_{\rm p}=\tilde{n}$, we obtain 
the densities $\tilde{n}_{\gamma}$ and $\tilde{n}_{+}$ in terms of 
$\tilde{n}$ and $n_{\rm p}$. When $\zeta \neq 1$ 
we obtain a quadratic equation for
$\tilde{n}_{+}$. It turns out that only one of its roots is
physical (i.e., both $\tilde{n}_{\gamma}$ and $\tilde{n}_{+}$ are 
non-negative). The physical root is given by
\beq
\tilde{n}_{+}=\frac{1}{2}\left[(\tilde{n}-n_{*})(1-\zeta^{2})^{-1}-n_{\rm p}
\right], \label{5.3.9}
\eeq
where $n_{*}=\zeta \surd[\tilde{n}^{2}-(1-\zeta^{2})n_{\rm p}^{2}]$.
When $\zeta = 1$ (which is true when $\widetilde{\Theta}=0.493$) we get
$\tilde{n}_{+}=(\tilde{n}-n_{\rm p})^{2}/(4\tilde{n})$. 
Therefore we 
have the necessary densities in terms of the temperature. When there are no 
ions ($n_{\rm p}=0$) these solutions take a simple form given by
\beq
\tilde{n}_{-}=\tilde{n}_{+}=\frac{K_{2}(1/\widetilde{\Theta})}{2\left[
\widetilde{\Theta}^{2}+K_{2}(1/\widetilde{\Theta})\right]}\tilde{n}
\quad{\rm and}\quad \tilde{n}_{\gamma} = \frac{\widetilde{\Theta}^{2}}{
\widetilde{\Theta}^{2}+K_{2}(1/\widetilde{\Theta})}\tilde{n}.
\label{5.3.10}
\eeq
Now we determine the temperature in terms of the initial data. We have
\beq
\tilde{u}_{\gamma}=\int_{0}^{\infty}4\pi \ee^{3}f_{\gamma}(\ee )d\ee=
3\widetilde{\Theta}\tilde{n}_{\gamma}   \label{5.3.11}
\eeq
and 
\beq
\tilde{u}_{\pm} = \int_{1}^{\infty}\gamma^{2}\sqrt{\gamma^{2}-1}f_{\pm}
(\gamma)d\gamma=\frac{3\widetilde{\Theta}K_{2}(1/\widetilde{\Theta})+K_{1}
(1/\widetilde{\Theta})}{K_{2}(1/\widetilde{\Theta})}\tilde{n}_{\pm}.
\label{5.3.12}
\eeq
Using the energy
conservation equation  $\tilde{u} = \tilde{u}_{\gamma}+
\tilde{u}_{-}+\tilde{u}_{+}$, we get the temperature as an implicit
function of $\tilde{u}$, $\tilde{n}$, and $n_{\rm p}$. In the limit
where $n_{\rm p}=0$, we have
\beq
\tilde{u}_{\gamma}=\frac{3\widetilde{\Theta}^{3}}{\widetilde{\Theta}^{2}+
K_{2}(1/\widetilde{\Theta})}\tilde{n}\quad {\rm and}\quad
\tilde{u}_{-}+\tilde{u}_{+}=\frac{3\widetilde{\Theta}K_{2}(1/\widetilde
{\Theta})+K_{1}(1/\widetilde{\Theta})}{\widetilde{\Theta}^{2}+
K_{2}(1/\widetilde{\Theta})}\tilde{n}
\label{5.3.13}.
\eeq
In this case, the equation for the temperature takes the form
\beq
3\widetilde{\Theta}^{3}+3\widetilde{\Theta}K_{2}(1/\widetilde{\Theta})+
K_{1}(1/\widetilde{\Theta})=\overline{\ee} \left[\widetilde{\Theta}^{2}
+K_{2}(1/\widetilde{\Theta})\right], \label{5.3.14}
\eeq
where $\overline{\ee}$ is the mean energy per particle (which is determined
by the initial conditions).
\mysubsec{6.2 thermal equilibrium:  densities and the temperature}

Here we determine the final temperature and  densities
resulting from  the radiative processes in the second phase. We have
$\mu_{-}+\mu_{+}=2\mu_{\gamma}=0$. Let $\mu_{+}=-\mu_{-}=\mu_{0}$ and 
$z = {\rm exp}(\mu_{0}/\Theta_{0})$. Clearly
\beq
n_{\pm}=\frac{8\pi}{\lambda_{0}^{3}}\Theta_{0}K_{2}(1/\Theta_{0})
z^{\pm 1}\quad {\rm and}\quad n_{\gamma}=\frac{16\pi \Theta_{0}^{3}}
{\lambda_{0}^{3}}. \label{5.3.15}
\eeq
Using the fact that $n_{-}=n_{\rm p}+n_{+}$ we can show that
\beq
n_{-}+n_{+} = \frac{16\pi}{\lambda_{0}^{3}}\Theta_{0}K_{2}(1/\Theta_{0})
\sqrt{1+x^{2}}, \label{5.3.16}
\eeq
where $x=\lambda_{0}^{3}n_{\rm p}/\left[16\pi \Theta_{0}K_{2}(1/\Theta_{0})
\right]$. In the nonrelativistic limit ($\Theta_{0}\ll 1$) the pair
density is given by
\beq
n_{-}+n_{+} = \frac{4}{\lambda_{0}^{3}}(2\pi\Theta_{0})^{3/2}\,{\rm exp}
(-1/\Theta_{0})\left[1+\frac{15}{8}\Theta_{0}+\frac{105}{128}\Theta_{0}^{2}
\right]\sqrt{1+x^{2}}.\label{5.3.16a}
\eeq
It can be shown that the pair energy density is given
by
\beq
u_{-}+u_{+}=\frac{16\pi}{\lambda_{0}^{3}}\left[3\Theta_{0}^{2}K_{2}(1/
\Theta_{0})+\Theta_{0}K_{1}(1/\Theta_{0})\right]\sqrt{1+x^{2}}.
\label{5.3.17}
\eeq
Finally, energy conservation implies
\beq
\frac{8\pi^{5}\Theta_{0}^{4}}{15}+16\pi\left[3\Theta_{0}^{2}K_{2}(1/
\Theta_{0})+\Theta_{0}K_{1}(1/\Theta_{0})\right]\sqrt{1+x^{2}}=
\lambda_{0}^{3}\tilde{u},   \label{5.3.18}
\eeq
where the first term on the left-hand side is the contribution
from the photons. 
We can solve this equation for $\Theta_{0}$ in terms
of $\tilde{u}$ and $x$ (equivalently $n_{\rm p}$). This completes the
analytical description of the thermal equilibrium state in terms of the
initial data. This treatment is exact and is valid for all energies
(relativistic or otherwise) and densities (so long as the plasma
is nondegenerate).
\mysubsec{6.3 time evolution of the spectra: two examples}

Now we consider the time evolution of the plasma for two specific 
initial conditions.
In the first case the initial photon and the pair distributions
are flat (i.e., $F$ is constant) and nonzero within the energy (in MeV) 
interval $0.1\leq \ee mc^{2}\leq 10$
and $0.1\leq (\gamma -1)mc^{2}\leq 10$. The initial densities are taken to
be $\nph = n_{+}+n{-} = 2\times 10^{20}\; {\rm cm}^{-3}$. For this case
we find a kinetic-equilibrium temperature  $\widetilde{\Theta}=
3.43$ and the corresponding densities are found to be  $\tilde{n}_{\rm ph}
= 1.36\times 10^{20}\;{\rm cm}^{3}$ and $\tilde{n}_{-}=\tilde{n}_{+}=1.32
\times 10^{20}\;{\rm cm}^{3}$. Monte Carlo evolution of the spectra for this
case are shown in figures $4$ and $5$.
\begin{figure}[p]
\centerline{\psfig{figure=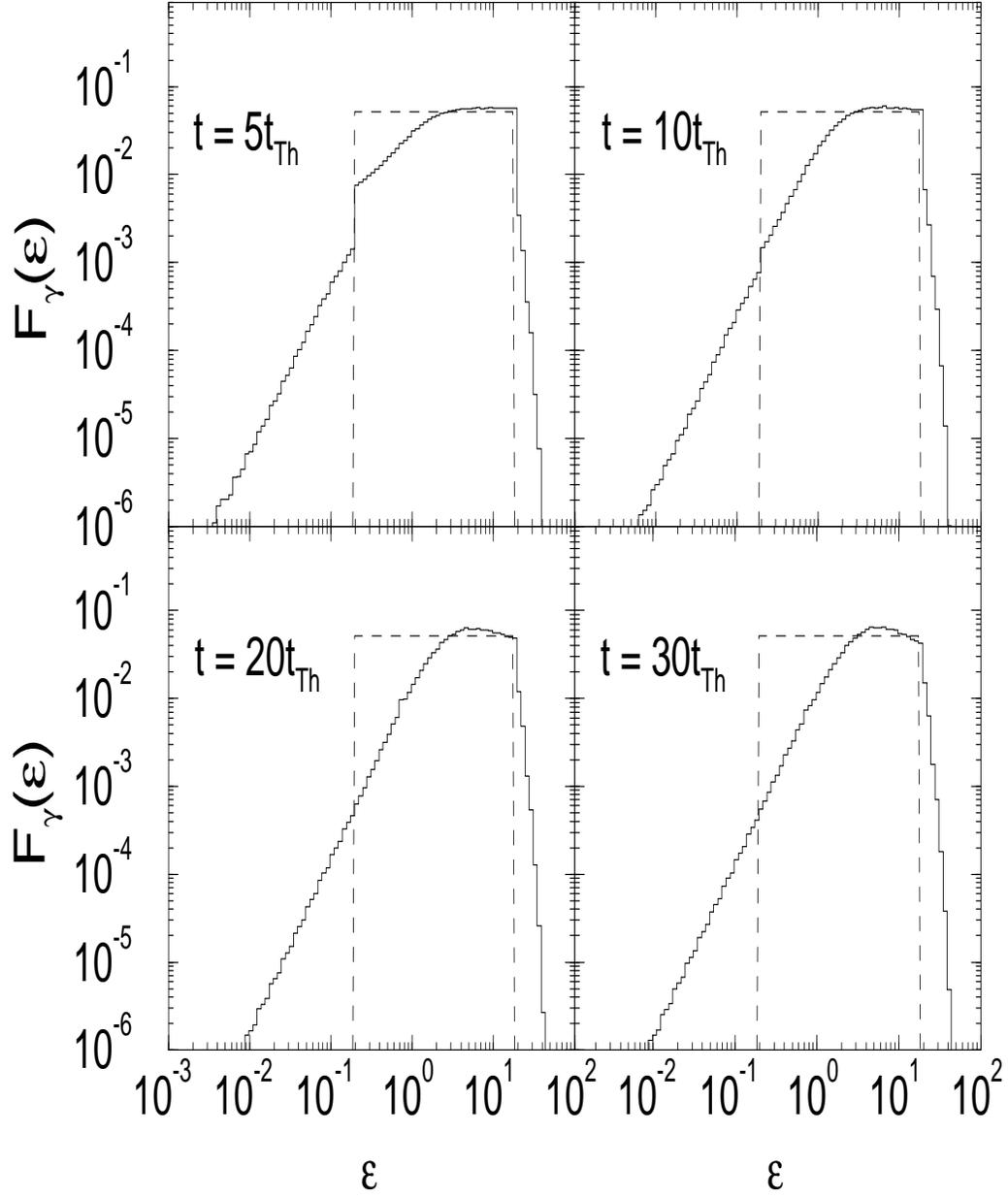,height=6.5in,width=5.5in}}
\caption{Time evolution of the photon spectrum (solid line) starting from a
flat initial spectrum (dashed line). Initial pair spectrum is 
flat as well. It is clear that the softer end of the spectrum relaxes first.
The same phenomenon is observed in the pair distribution (not shown here).}
\end{figure}
\begin{figure}[p]
\centerline{\psfig{figure=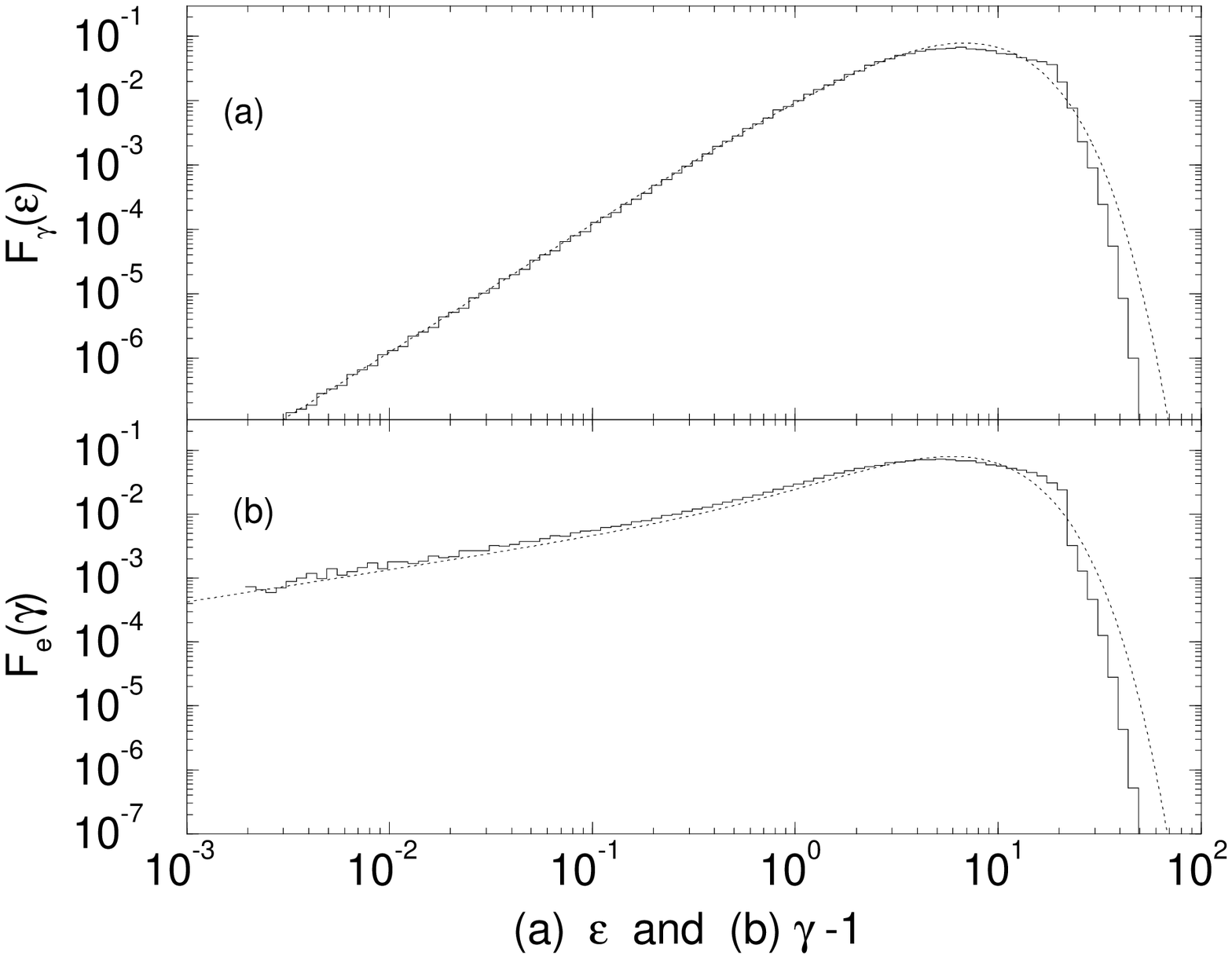,height=6.5in,width=5.5in}}
\caption{Final Monte Carlo spectra (the solid histograms) at $t=45 t_{\rm Th}$
compared with the analytical solution (dashed curves) for
(a) the photons and (b) the pairs, starting from the flat initial 
spectra. See section $6.3$ for details.}
\end{figure}
They agree well with the analytical 
kinetic-equilibrium solutions. For this case, as well as the second one, 
we have used $t_{\rm Th} = (c\sigma_{\rm Th}n)^{-1}$, with $n = 2\times 10^{20}
\;{\rm cm}^{3}$. In the second case we start with the same densities of
the photons and pairs and  the initial distributions are confined to the 
same band width as above. The only difference is that $F_{\gamma}(\ee)\propto
\ee^{-2}$  and $F_{\rm e}(\gamma)\propto \gamma^{-2}$, with suitable
normalizations. In this case we obtain a kinetic-equilibrium 
temperature $\widetilde{\Theta}=0.663$ and the corresponding densities are
found to be  $\tilde{n}_{\rm ph}=1.73\times 10^{20}\; {\rm cm}^{-3}$ and
$\tilde{n}_{-}=\tilde{n}_{+}=1.1\times 10^{20} \;{\rm cm}^{-3}$.
Monte Carlo spectra for this case are shown in figures $6$ and $7$. 
\begin{figure}[p]
\centerline{\psfig{figure=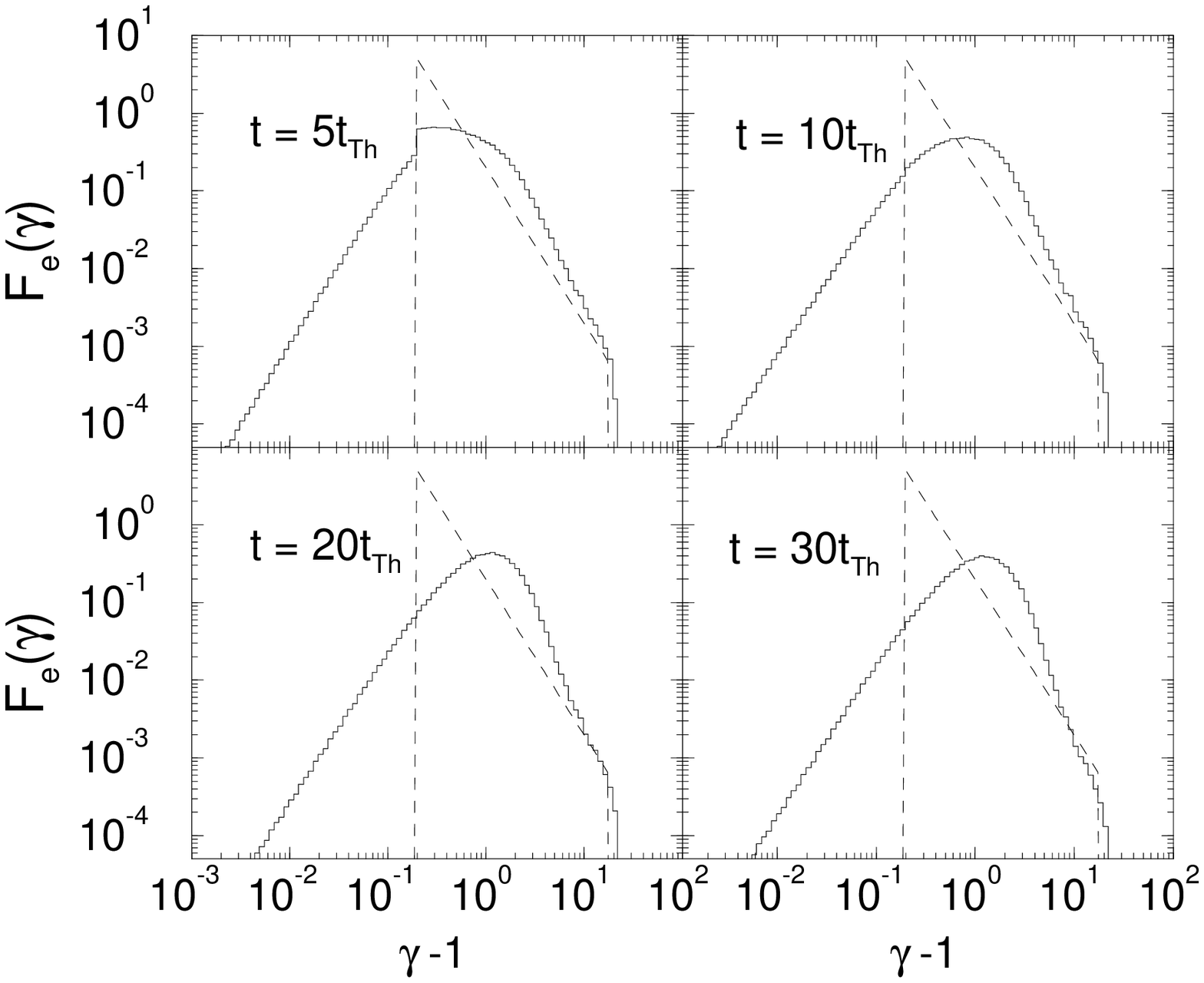,height=6.5in,width=5.5in}}
\caption{Evolution of the photon spectrum (solid line) starting from the
power law ($\delta = 2$) distributions of the photons 
(dashed line) and the pairs
(evolution not shown here). As in the previous example, the relaxation is 
faster at lower energies. }
\end{figure}
\begin{figure}[p]
\centerline{\psfig{figure=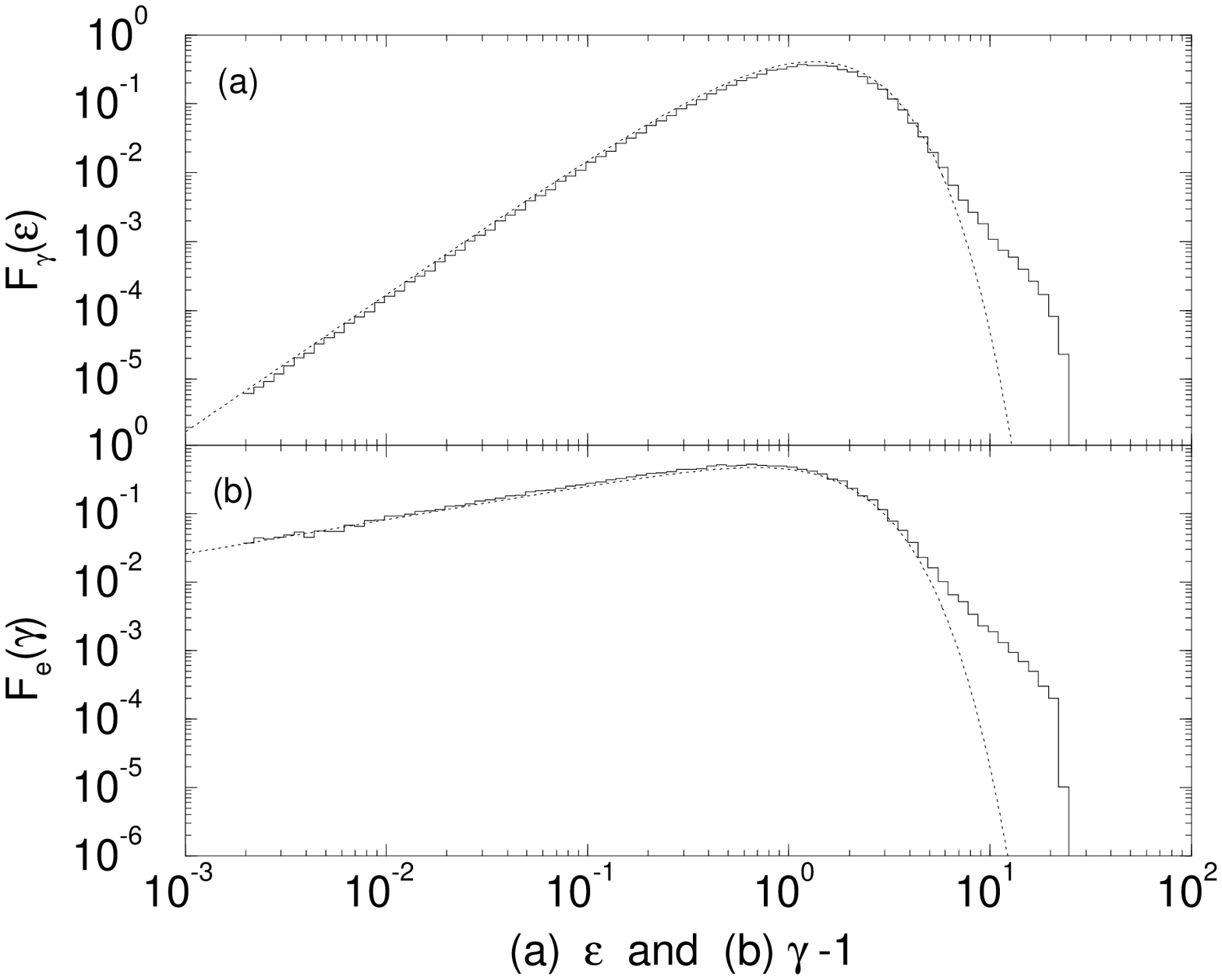,height=6.5in,width=5.5in}}
\caption{Final Monte Carlo spectra (the solid histograms) at $t=40 t_{\rm Th}$
compared with the analytical solution (dashed curves) for
(a) the photons and (b) the pairs, starting from the power law initial 
spectra. It is evident that the high-energy tails persist for a long
time, becoming steeper with time
(analogous to the relaxation in a non-relativistic plasma), 
but the number of particles 
(and the energy) in these tails is less than a few percent of the total.}
\end{figure}
Once again
they are in a good agreement with the analytical solution.
We have verified the number and energy conservation 
after each time-step. 
The final densities are found to agree with the predicted values 
within an accuracy of $10\%$ or better (which can be improved by using more
energy bins). In both cases the kinetic-equilibrium 
solution is moderately relativistic.
It is clear from  figures $5$ and $7$ that the low-energy part of the 
spectrum relaxes before the high-energy end. 
The cross sections (and hence the reaction rates) decrease with the energy,
thereby making the relaxation slower at higher energies. 
A part of the deviation from the analytical solutions that we see
in  figures $5$ and $7$ (in the high-energy tails)
could be an artifact of our sparse (logarithmic)
binning at higher energies. It can be rectified by using more 
bins in the high-energy end (more computing time).
For the above cases the final thermal-equilibrium
temperatures turn out to be $\Theta_{0}=
4.36 \times 10^{-3}$ and $\Theta_{0}=2.98\times 10^{-3}$,
respectively. 
\medskip
%
%
%
%
%
%
%
\mysection{7 Conclusions}
\reset
\smallskip

We have developed a new computational method for 
solving the Boltzmann equations of a pair plasma which is applicable
for arbitrary energies (in the X-ray and $\gamma$-ray bands),
densities, and distribution functions. 
We have fully analyzed all relevant microscopic processes in a 
pair plasma viz., Comptonizaion, the pair creation and annihilation, 
bremsstrahlung and the associated cooling, and Coulomb collisions.
The spectra from the individual collision integrals, using our expressions
and the numerical method (for Compton scattering, pair annihilation, and 
bremsstrahlung), are in a good agreement with several previous results 
obtained by using  different methods (e.g., S82a; CB90; and Dermer 1986). 
The analysis given in this paper can be 
very easily extended to an inhomogeneous and anisotropic plasma.
It will only change some of the collision integrals and add a 
spatial component to the kinetic equations. That will result in an increase
in the computational time but it will still be  
manageable by the present day work stations.
Presence of the magnetic fields will alter the kinetics
(through synchrotron emission) and it can be modeled along the same lines
as that of C92. We have developed a modified version
of the adaptive Monte Carlo method which is very efficient 
and robust. It is faster than the crude Monte Carlo method 
(using uniform sampling) by at least a factor of ten and is more flexible
than the  numerical integration methods (which do not use random
sampling) which are used in the past.
We have obtained the analytical equilibrium solutions for a general set 
of initial conditions.
Finally we have tested our Monte Carlo evolution scheme for two 
specific sets of initial conditions and found that the results compared 
favorably with the corresponding analytical solutions. The method
is found to be very stable. In each of the examples considered,
the program has analyzed a total of $\sim 10^{10}$ collision events.
This stability, accompanied by its generality and the inherent
flexibility, makes this technique suitable for many astrophysical
applications. In particular, this formalism can be applied to the 
expanding pair plasmas in the $\gamma$-ray-burst sources  in their 
final stages of evolution (when they are only moderately optically 
thin), AGN, and the emission from hot accretion discs near black holes.

 RP is very pleased to thank Marc Kamionkowski for numerous discussions
and insightful suggestions. He is happy to thank Malvin Ruderman
for many discussions, help, and encouragement. He thanks Alexandre 
Refregier and Abraham Loeb for interesting discussions and Andrzej 
Zdziarski for sending his papers on this subject. Thanks are due
to the referee, Paolo Coppi, for many useful suggestions which 
made this paper much more concise and clear. This work was 
partially supported by a grant from NASA, NAG 5-2841.
\medskip
%
%
%
%
%
%
%
~~~\\
\centerline{\bf Appendix}
\reset
\def\theequation{{A.\arabic{equation}}}
\parindent=1cm
\vspace{-14pt}
\mysubsec{Compton scattering rate for photons}

The cross section in the C-frame is given by
\beq
\frac{d\sigma}{dP}=\frac{1}{\ee^{2}}\frac{d\sigma}{d\Omega}\delta(\ee-\tilde
{\ee}),
\label{main cross section}
\eeq
with $\delta(\ee - \tilde{\ee})=\delta(\ee_{1}-\tilde{\ee}_{1})
d\tilde{\ee}_{1}/d\ee$.  Here  $d\Omega$ is an infinitesimal 
solid angle around 
the direction $\bf k$ (similarly $d\Omega^{\prime}$ is defined with respect to
${\bf k}^{\prime}$).  It is easy to see that 
$d\tilde{\ee}_{1}/d\ee=\xi^{2}\,a\,a_{1}^{-1}$. Now
\beq
\left(\frac{d\sigma}{d\Omega}\right)_{\rm C-frame}=
\mbox{\hspace{.5cm}}\frac{d\Omega^{\prime}}{d\Omega}
\left(\frac{d\sigma}{d\Omega}\right)_{\rm R-frame}.
\label{cross section}
\eeq
Since $\ee^{2}d\Omega=\ee^{\prime\,2}d\Omega^{\prime}$ we get
$d\Omega/d\Omega^{\prime}=(\gamma a)^{2}$. Finally,
\beq
\left(\frac{d\sigma}{d\Omega}\right)_{\rm R-frame}=\mbox{\hspace{.5cm}}
\frac{\re^{2}\Delta}{2\,\xi^{3}}
\label{Klein-Nishina}
\eeq
is the Klein-Nishina formula in our notation, where 
$\Delta=\xi^{2}-\xi\,\sin^{2}\theta^{\prime}+1$. This leads to 
\beq
\frac{d\sigma}{dP}=\frac{\re^{2}}{\ee^{2}a_{1}}\,\frac{\Delta}
{2\,\gamma^{2}a\,\xi}\,\delta(\ee_{1}-\tilde{\ee}_{1}).
\label{final cross section}
\eeq
In equation (\ref{final-eta}) we set $\brel=1$, $\fkin_{12}=a_{1}$,
$n_{1}=\nph$, $n_{2}=n_{-}+n_{+}$, $F_{1}=F_{\gamma}$,
$F_{2}=F_{e}$, $\ee_{2}=\gamma$, and 
$\delta_{12}=0$. Clearly $d\Omega_{1}\,d\Omega_{2}=2\,\pi d\mu\,d\mu^{\prime}
\,d\phi$, $a=1-\beta\,\mu^{\prime}$, 
$a_{1}=1-\beta\,\mu^{\prime \prime}$, where
$\mu^{\prime \prime}=
\mu\,\mu^{\prime}+\surd[(1-\mu^{2})(1-\mu^{\prime\,2})]
\cos{\phi}$, and
$b=1-\mu$. These substitutions lead to equation (\ref{compton eta}).
\smallskip
\mysubsec{Bremsstrahlung emissivity}

Here we derive equation (\ref{brems pair eta}) and explain the notation
used in that connection.
We are interested in the processes in which two particles of momenta $p_{i}=
\gamma_{i}(1,\bb_{i}),\:i=1,2$ radiatively scatter on each other to produce a 
photon of momentum $p=\ee(1,{\bf k})$. Here $c\bb_{i}$ are the particle 
velocities  in the C-frame and $\gamma_{i}$
are the corresponding Lorentz factors, $\ee$ is the energy of the emitted
photon, and $\bf k$ is its directional unit vector. 
Let $\mu$, $\mu^{\prime}$, and $\mu^{\prime \prime}$ be the cosines of 
the angles between the pairs of vectors $(\bb_{1},\bb_{2})$, $(\bb_{1},{\bf 
k})$, and $(\bb_{2},{\bf k})$, respectively.
The angle between the planes formed by the pairs of vectors
$(\bb_{1},\bb_{2})$ and $(\bb_{1},{\bf k})$
is defined to be $\phi$. We have $\mu^{\prime \prime}=\mu\,\mu^{\prime}
+\surd[(1-\mu^{2})(1-\mu^{\prime\,2})]\cos{\phi}$. 
In equation (\ref{final-eta}), because of the isotropy of the distribution 
functions, we can write $d\Omega_{1}d\Omega_{2}=2\,\pi\,d\mu\,d\Omega$,
where $d\Omega$ is an infinitesimal solid angle around $\bf k$.
We define $(d\sigma/d\ee)_{i}=\ee^{2}\int\,d\Omega(d\sigma/dP)_{i}$. The
case $i=1$ refers to the $e^{\pm}$-$e^{\pm}$ process and the case $i=2$ 
refers to the $e^{\pm}$-$e^{\mp}$ bremsstrahlung. 
It is shown by Haug (1975b) that
\begin{eqnarray}
\left(\frac{d\sigma}{d\ee}\right)_{i} &=& \frac{\alpha\re^{2}\ee}{\pi}
\int d\Omega\,\frac{C_{i}}{\rho\Delta_{i}},\mbox{ if }\ee\leq\ee^{*},
\nonumber\\&&\nonumber\\[-10pt]
 &=&0,\mbox{ otherwise,}
\end{eqnarray}
where $\Delta_{1}=\omega\sqrt{\omega^{2}-4}$, 
$\Delta_{2}=2\sqrt{\zeta^{2}-1}$, and 
$\rho=\surd{[\omega^{2}-2(x_{1}+x_{2})]}$, while 
$\omega=\surd{[2\,(\zeta + 1)]}$, 
$\zeta=p_{1}\cdot p_{2}=\gamma_{1}\gamma_{2}(1-\beta_{1}
\,\beta_{2}\,\mu)$, $x_{1}=p\cdot p_{1}=\ee\gamma_{1}(1-\beta_{1}
\,\mu^{\prime})$,
and $x_{2}=p\cdot p_{2}=\ee\gamma_{2}(1-\beta_{2}\,\mu^{\prime \prime})$.
Here $\alpha$ is the fine-structure constant.
Finally
\beq
\ee^{*}=\frac{\zeta - 1}{\gamma_{1}+\gamma_{2}-\sqrt{(\gamma_{1}+
\gamma_{2})^{2}-2(\zeta +1)}},
\eeq
and 
\beq
C_{i}=\left\{\frac{\sqrt{\rho^{2}-4}}{\pi}\int\,Ad\Omega^{\prime}\right\}_{i}.
\eeq
The cross section
$C_{1}$ was computed by Haug (1975a, eq.[A1]) and $C_{2}$ 
by Haug (1985a, eq.[A1]). This latter cross section has
some minor errors and the corrections are given in Haug (1985b). For
$C_{1,2}$ we have followed the notation of Haug except that $d\Omega^{\prime}$
was called $d\Omega_{p_{1}^{\prime}}$ in $C_{1}$ and it was called 
$d\Omega_{q^{\prime}}$ in $C_{2}$, in those papers.
Going back to equation (\ref{final-eta}) we 
have to set $\delta_{12}=1$ for the case
$i=1$. Hence $n_{1}n_{2}\rightarrow \frac{1}{2} (
n_{+}^{2}+n_{-}^{2})$. In the second case $\delta_{12}=0$ and 
$n_{1}n_{2}\rightarrow n_{+}n_{-}$. We have  $F_{1}=F_{2}=F_{e}$ and 
$\ee_{i}=\gamma_{i}$ for $i=1,2$. With these substitutions the desired result 
follows. In the present notation $\fkin_{12}=\sqrt{\zeta^{2}-1}/
\gamma_{1}\,\gamma_{2}$ and $d\Omega=d\mu^{\prime}\,d\phi$. 
The integration domain $U$ is specified by $\gamma_{\rm min}\leq\gamma_{1,2}
\leq\gamma_{\rm max}$, $-1\leq\mu,\,\mu^{\prime}\leq 1$, and $0\leq\phi
\leq 2\,\pi$, subject to the condition that $\ee^{*}(\gamma_{1},\,
\gamma_{2},\,\mu)\geq\ee$.
\smallskip
\mysubsec{Compton scattering rate for pairs}

The cross section can be written as (see e.g., JR80)
\beq
\sigma=\frac{\re^{2}}{2\,\ee\,\gamma\,\rho_{1}}
\int\,d\tau_{f}\,\delta^{(4)}(q+q_{1}-p-p_{1})\,X,
\label{e-total-cross}
\eeq
where $d\tau_{f}=d^{3}{\bf q}\,d^{3}{\bf q}_{1}$,
${\bf q}=\gamma\,\bb$,
${\bf q}_{1}=\ee\,{\bf k}$, while $X$ is given by equation (\ref{e-x}).
Here $\rho_{1}=p\cdot p_{1}=q\cdot q_{1}$ and $\rho_{2}=p\cdot q_{1}=
p_{1}\cdot q$. In equation (\ref{e-total-cross}) we can remove three of the
delta functions by integrating over $d^{3}{\bf q}^{\prime}$.
Using the conservation of three-momentum we obtain
$\ee^{2}=(\gamma_{1}\bb_{1}+\ee_{1}{\bf k}_{1}-\gamma \bb)^{2}$ and
$d\ee/d\gamma=(\gamma \beta -\gamma_{1} \beta_{1}
\mu^{\prime}-\ee_{1}\mu)/\beta \ee$, while $\mu$ and $\mu^{\prime}$ are 
defined in section $4$. After some straight forward manipulations we find
\beq
\frac{d\sigma}{dP}=\frac{\re^{2}}{2\gamma \ee \rho_{1}}
\left|\frac{d\tilde{\ee}_{1}/d\gamma}{1+d\ee/d\gamma}\right| 
X \delta(\ee_{1}-\tilde{\ee_{1}}),
\label{electron-comp-cross}
\eeq
where $dP=d^{3}{\bf q}=\beta\,\gamma^{2}\,d\gamma\,
d\Omega$, and  $d\Omega$ is the infinitesimal solid angle around the
direction $\bb$. Now, in equation (\ref{final-eta}) we set $\beta_{\rm rel}=1$,
$\fkin_{12}=a_{1}$, $n_{1}=n_{\gamma}$, $n_{2}=n_{-}+n_{+}$, 
$F_{1}=F_{\gamma}$, $F_{2}=F_{e}$,
$\delta_{12}=0$, and $\ee_{2}=\gamma_{1}$.
Clearly $a=1-\beta\,\mu$, 
$a_{1}=1-\beta_{1}\,\mu^{\prime \prime}$, where $\mu^{\prime \prime}=
\mu\,\mu^{\prime}
 +\surd{[(1-\mu^{2})(1-\mu^{\prime 2})]}\cos{\phi}$, and 
$b=1-\beta\,\beta_{1}\,\mu^{\prime}$. Finally, $d\Omega_{1}\,d\Omega_{2}=
2\,\pi\,d\mu\,d\mu^{\prime}\,d\phi$. These substitutions lead to 
equation (\ref{e-comp-emis}).
\medskip
\mysubsec{Pair creation rate}
\medskip
Let $d\Omega_{i}$ be the infinitesimal solid 
angles around ${\bf k}_{i}$ for $i=1,2$. Infinitesimal solid angles around
$\bb$ and $\bb_{\rm cm}$ are denoted by $d\Omega$ and $d\Omega_{\rm cm}$,
respectively. In equation (\ref{final-eta}) we have $d\Omega_{1}d\Omega_{2}=
2\pi\,d\mu d\Omega$, because of the isotropy. We define $d\sigma/d\gamma
=\beta\gamma^{2}\int d\Omega(d\sigma/dP)$,
where $dP=\beta\gamma^{2}d\gamma d\Omega$. We set $n_{1}=n_{2}=\nph$,
$\delta_{12}=1$, $\fkin_{12}=1-\mu$, and $F_{1}=F_{2}=F_{\gamma}$.
It can be shown that
\beq
\frac{d\sigma}{d\gamma}=\int d\Omega_{\rm cm}\frac{d\gamma_{\rm cm}}{d\gamma}
\left( \frac{d^{2}\sigma}{d\gamma d\Omega}\right)_{\rm cm}
=\int d\Omega_{\rm cm}
\frac{d\gamma_{\rm cm}}{d\gamma}\left(\frac{d\sigma}{d\Omega}\right)_
{\rm cm}H(\ee_{\rm cm})\,\delta(\gamma_{\rm cm}-\ee_{\rm cm}),
\label{pair-cr-b}
\eeq
where $H$ is the Heaviside step function which is zero for negative
arguments and is unity otherwise. The latter imposes the 
pair creation threshold. It can be easily seen that
$d\Omega_{\rm cm}=dz\,d\phi$. The delta function in the last 
equation ensures energy conservation. It can be written in the form
$\delta(\gamma_{\rm cm}-\ee_{\rm cm})=
\left|d\tilde{z}/d\gamma_{\rm cm}\right|\delta(z-\tilde{z})$, 
where $\tilde{z}=(\gamma_{\rm c}\gamma_{\rm cm}-\gamma)\Delta^{-1}$ 
and $\Delta=\beta_{\rm c}\beta_{\rm cm}\gamma_{\rm c}\gamma_{\rm cm}$.
This is the solution to the equation $\gamma=\gamma_{\rm c}\gamma_{\rm cm}
(1-\beta_{\rm c}\beta_{\rm cm}z)$ 
(i.e., $p\cdot q=p_{\rm cm}\cdot q_{\rm cm}$). Finally,
$\left|d\tilde{z}/d\gamma\right|=\Delta^{-1}$. After all these 
substitutions in equation (\ref{final-eta}) we arrive at 
equation (\ref{pair-cr-f}).
\medskip
\mysubsec{bremsstrahlung cooling functions}
\medskip

Here we give the cooling functions used in equation(\ref{cool-b}).
The energy radiated per unit time in $e^{\pm}$-proton collisions is given
by 
\beq
E_{ep}(\gamma)=c n_{\rm p}
\int_{0}^{\gamma-1}\,d\ee\,\ee\left(\frac{d\sigma}{d\ee}
\right)_{\rm proton},
\eeq
where the protons are assumed to be at rest. Here $\ee$ is the
energy of the emitted photon and $d\sigma/d\ee$ is the cross section 
(see e.g., JR80). For $E_{ee}$ we
start from equation (3.5) of Haug (1975b). After some algebra we arrive at 
\beq
E_{ee}(\gamma,\gamma^{\prime})=\frac{c(n_{+}^{2}+n_{-}^{2})}{2\ne}\,
\frac{\gamma+\gamma^{\prime}}{\gamma\gamma^{\prime}}\,
\int_{-1}^{1}\,d\mu\,p_{c}\,Q_{ee}(\ee_{c},p_{c}),
\eeq
where $\ee_{c}=\surd{[(\zeta+1)/2]}$, 
 $p_{c}=\surd{[(\zeta-1)/2]}$, and $\zeta=\gamma\gamma^{\prime}(
1-\beta\beta^{\prime}\mu)$, while  $\mu$ is the cosine of the interaction
angle. Averaging over this angle ($\mu$-integration) gave rise to
the factor of half above.
Presence of $\ne$ in the denominator 
is a consequence  of our definition of $E_{ee}$. The
cooling function $Q_{ee}$, which is accurate to $\sim 6\%$ or better, 
is given by equation (3.15) of Haug (1975b). We reproduce it here for 
convenience: 
\beq
Q_{ee}\approx 8\alpha\re^{2}\,\frac{p_{c}^{2}}{\ee_{c}}\,\left[
1-\frac{4}{3}\,\frac{p_{c}}{\ee_{c}}+\frac{2}{3}\,\left(2+\frac{p_{c}^{2}}
{\ee_{c}^{2}}\right)\ln{(\ee_{c}+p_{c})}\right],
\eeq
where $\alpha$ is the fine-structure constant. For $e^{\pm}$-$e^{\mp}$
process we get
\beq
E_{e\bar{e}}(\gamma,\gamma^{\prime})=\frac{cn_{+}n_{-}}{2\ne}\,
\frac{\gamma+\gamma^{\prime}}{\gamma\gamma^{\prime}}\,
\int_{-1}^{1}\,d\mu\,p_{c}\,Q_{e\bar{e}}(\ee_{c},p_{c}).
\eeq
The cooling function $Q_{e\bar{e}}$ is given by equations
(26) and (28) of Haug (1985c). 
For the sake of convenience, we reproduce it here:
\beq
Q_{e\bar{e}}=\left\{ \begin{array}{ll}
\frac{32}{3}\alpha\re^{2}\,\sum_{i=0}^{4}a_{i}p_{c}^{i}&\mbox{if $E_{c}\le300
$ KeV,}\\[5pt]
16\alpha\re^{2}\left(\ee_{c}\ln{(\ee_{c}+p_{c})}-
\frac{1}{6}\,\ee_{c}+\sum_{i=0}^{2}\,b_{i}\ee_{c}^{-i}\right)
&\mbox{otherwise,}\end{array}
\right.
\eeq
where $a_{0}=1.096$, $a_{1}=-0.523$, $a_{2}=0.1436$, $a_{3}=1.365$,
$a_{4}=-0.532$, $b_{0}=-0.726$, $b_{1}=1.575$, and $b_{2}=-0.796$.
Here $E_{c}=mc^{2}\ee_{c}$.
\smallskip
\mysubsec{Landau collision integral for Coulomb collisions}

The flux vector (see Lifshitz \& Pitaevskii 1981 -- LP81 henceforth) is
given by
\beq
S_{1}^{i}({\bf p})=\sum_{s=1}^{2}\int d^{3}{\bf p}^{\prime}\left(
f_{1}({\bf p})\frac{\partial}{\partial p^{\prime j}}f_{s}({\bf p}^{\prime})-
f_{s}({\bf p}^{\prime})\frac{\partial}{\partial p^{j}}f_{1}
({\bf p})\right)\,B^{ij}
.
\label{land-b}
\eeq
The superscripts $i$, $j$ 
in this equation denote the components of three-vectors or tensors. 
In equation (\ref{land-b}) the 
summation over $j$ is implicit. The components of
momenta are given by $p^{i}=\gamma\beta^{i}$ and $p^{\prime i}=\gamma^{\prime}
\beta^{\prime i}$, for $i=1,2,3$. We have $d^{3}{\bf p}^{\prime}=\beta^{\prime}
\gamma^{\prime 2}d\Omega^{\prime}$. Let $\zeta=\gamma\gamma^{\prime}(1-\beta
\beta^{\prime}\mu)$, where $\mu$ is the cosine of the interaction angle. 
The tensor $B^{ij}$ (see LP81) is given by
\beq
B^{ij}=\frac{2\pi\,c\,\re^{2}\,
\ln{\Lambda_{\rm C}}\,\zeta^{2}}{\gamma\,\gamma^{\prime}
\,(\zeta^{2}-1)^{3/2}}\left[
(\zeta^{2}-1)\delta^{ij}-\beta^{i}\beta^{j}\gamma^{2}
-\beta^{\prime i}\beta^{\prime j}\gamma^{\prime 2}+(\beta^{i}\beta^
{\prime j}+\beta^{j}\beta^{\prime i})\gamma\gamma^{\prime}\zeta\right],
\label{land-c}
\eeq
where we have made some slight modifications to take into account the 
dimensions of the distributions and the momenta. 
This tensor satisfies the identity
$\sum_{j=1}^{3}B^{ij}(\beta^{j}-\beta^{\prime j})=0,\,\,{\rm for}\,\, i=1,2,3.$
For isotropic distributions we have $\partial f_{s}/\partial p^{j}=\beta^{j}\,
\partial f_{s}/\partial \gamma$. Using this fact and the previous 
identity we obtain
\beq
S_{1}^{i}=\int d\gamma^{\prime}d\Omega^{\prime}\beta^{\prime}\gamma^{\prime 2}
\,{\cal D}_{1}(\gamma,\gamma^{\prime})\sum_{j=1}^{3}B^{ij}\beta^{j},
\eeq
where 
\beq
{\cal D}_{1}(\gamma,\gamma^{\prime})
=\sum_{s=1}^{2}\left[f_{1}(\gamma)\frac{\partial}{\partial 
\gamma^{\prime}}f_{s}(\gamma^{\prime})-f_{s}(\gamma^{\prime})
\frac{\partial}{\partial \gamma}f_{1}(\gamma)\right].
\eeq
We choose a coordinate frame in which $\beta^{1}=\beta$,
$\beta^{2,3}=0$, $\beta^{\prime 1}=\beta^{\prime}\,\mu$, $\beta^{\prime 2}
=\beta^{\prime}\sqrt{1-\mu^{2}}$, and $\beta^{\prime 3}=0$.
Also $d\Omega^{\prime}=2\pi d\mu$.
With these substitutions we find
\beq
\frac{\partial}{\partial t}f_{1}(\gamma)=-\beta\frac{\partial}{\partial \gamma}
\int\,\,2\pi d\mu \,d\gamma^{\prime}\beta\beta^{\prime}\gamma^{\prime 2}
B\,{\cal D}_{1}(\gamma,\gamma^{\prime}),
\label{land-g}
\eeq
where $B=2\pi\,c\,\re^{2}\,\ln{\Lambda_{\rm C}}\,B_{0}$ and
\beq
B_{0}=\frac{\zeta^{2}}{\gamma\gamma^{\prime}(\zeta^{2}-1)^{3/2}}\,
\left(\zeta^{2}-1-\beta^{2}\gamma^{2}-\beta^{\prime 2}\gamma^{\prime 2}
\mu^{2}+2\beta\beta^{\prime}\gamma\gamma^{\prime}\mu\zeta\right).
\label{land-h}
\eeq
The integral in equation (\ref{land-g}) is the Landau collision integral
for small angle deflections. This leads to the required result.
\eject
%
%
%
%
%
%
%
\centerline{\bf References}
\parindent=0pt
\parskip=0pt
\baselineskip=19pt
\def\aj#1#2{{Astrophys. J.}, {\bf #1}, #2}
\def\rs#1#2{{Mon. Not. R. Astr. Soc.}, {\bf #1}, #2} 
\vspace{-10pt}
\begin{tabbing}
Baring, M. \=1987a, \rs{228}{681}\\
           \>1987b, \rs{228}{695}\\
\end{tabbing}
\vspace{-35truept}
Bisnovatyi-Kogan, G. S., Zel'dovich, Ya. B., \& Sunyaev, R. A. 1971,\\
{Soviet Astr. -- AJ}, {\bf 15}, 17\\
Blumenthal, G. R. \& Gould, R. J. 1970, {Rev. Mod. Phy.}, {\bf 42}, 237\\
Chen, K. \& Ruderman, M. 1993, \aj {402}{264}\\
Coppi, P. S. 1992, \rs {258}{657} (C92)\\
Coppi, P. S. \& Blandford, R. D. 1990, \rs {245}{453} (CB90)\\
de Groot, S. R., van Leeuwen, W. A., \& van Weert, Ch. G., 1980,\\
{\it Relativistic Kinetic Theory} (Amsterdam: North-Holland)\\
Dermer, C. D. 1986, \aj{307}{47}\\ 
Dermer, C. D. \& Liang, E. P. 1989, \aj {339}{512}\\
Fabian, A. C., Blandford, R. D., Guilbert, P. W., Phinney, E. S., \& Cuellar,
L. 1986, \rs {221}{931}\\
Ghiselleni, G. 1987, \rs {224}{1}\\
Guilbert, P. W. \& Stepney, S. 1985, \rs {212}{523}\\
Gunnlaugur, B. \& Svensson, R. 1992, \aj {394}{500}\\
\vspace{-29truept}
\begin{tabbing}
Haug, E. \=1975a, {Z. Naturforschung}, {\bf 30a}, 1099\\
         \>1975b, {Z. Naturforschung}, {\bf 30a}, 1546\hfill\\
         \>1985a, {Phy. Rev. D}, {\bf 31}, 2120\hfill\\
         \>1985b, {Phy. Rev. D}, {\bf 32}, 1594(E)\hfill\\
         \>1985c, {Astron. Astrophys.}, {\bf 148}, 386\hfill\\
         \>1987, {Astron. Astrophys.}, {\bf 178}, 292\hfill\\
         \>1989, {Astron. Astrophys.}, {\bf 218}, 330\hfill\\ 
\end{tabbing}
\vspace{-29truept}
Jauch, J. M. \& Rohrlich, F. 1980, {\it The Theory of Photons and Electrons}\\
(Berlin: Springer-Verlag) (JR80)\\
Kusunose, M. 1987, \aj {321}{186}\\
Landau, L. D. \& Lifshitz, E. M. 1975, {\it Classical Theory of Fields}
(New York: Pergamon)\\
Lepage, P. 1978, {J. Comp. Phys.}, {\bf 27}, 192\\
Lifshitz, E. M. \& Pitaevskii, L. P. 1981, {\it Physical Kinetics} 
(New York: Pergamon) (LP81)\\
Lightman, A. P. 1981, \aj {244}{392}\\
--------------- 1982, \aj {253}{842}\\
Lightman, A. P. \& Band, D. L. 1981, \aj {251}{713}\\
Lightman, A. P. \& Zdziarski, A. A. 1987, \aj {319}{643}\\
M\'esz\'aros, P. {\&} Rees, M. J. 1993a, \aj {418}{L59}\\
--------------------------.1993b, \aj {405}{278}\\
Novikov, I. D. \& Stern, B. E. 1986 in {\it Structure and Evolution of 
Active Galactic Nuclei,} p. 149, edited by Giuricin, G., Mardirossian, F.,
Mezzetti, M., \& Ramella, M. (Reidel, Dordrecht)\\
Padovani, P., 1996, astro-ph/9610155\\
Piran, T. \& Shaham, J., 1977, {Phy. Rev. D}, {\bf 16}, 1615\hfill\\
Podznyakov, L. A., Sobol', I. M., \& Sunyaev, R. A. 1977, Soviet Astr.
{\bf 21}, 708\\
Press, W. H. et al. 1992 , {\it Numerical Recipes}
(New York: Cambridge University Press)\\
Ramaty, R., \& McKinley, J. M., \& Jones, F. C. 1982, \aj {256}{238}\\
Ramaty, R. \& M\'esz\'aros, P. 1981, \aj {250}{384}\\
Rybicki, G. B. \& Lightman, A. P. 1979, {\it Radiative Processes in 
Astrophysics},\\ chapter 7 (New York: John Wiley)\\
Sikora, M. 1994,  {Astrophys. J. Supp.}, {\bf 90}, 923\\
Stern, B. E. 1985, Soviet Astr. {\bf 29}, 306\\
Stern, B. E., Begelman, M. C., Sikora, M., \& Svensson, R. 1995,
\rs {272}{291}\\
Sunyaev, R. A., et al. 1992 \aj {389}{L75}\\
\vspace{-32truept}
\begin{tabbing}
Svensson, R. \=1982a, \aj {258}{321 (S82a)}\\
             \>1982b, \aj {258}{335 (S82b)}\\
             \>1987, \rs {227}{403}\\ 
             \>1994, {Astrophys. J. Supp.}, {\bf 92}, 585\\
\end{tabbing}
\vspace{-32truept}
Tanaka, F. \& Kusunose, M. 1985, {Prog. Th. Phy.}, {\bf 73}, 1390\\
Yahel, R. Z. \& Brinkmann, W. 1981, \aj{244}{L7}\\ 
\vspace{-32truept}
\begin{tabbing}
Zdziarski, A. A. \=1980, {Acta Astr.}, {\bf 30}, 371\\
                 \>1985, \aj {289}{514}\\
                 \>1988, \aj {335}{786}\\
	         \>1989, \aj {342}{1108}\\
\end{tabbing}
\vspace{-32pt}
Zdziarski, A. A., Coppi, P. S., \& Lamb, D. Q. 1990, \aj {357}{149}\\
\eject
\end{document}